\RequirePackage{snapshot}
\documentclass[aps,prb,twocolumn,amsfonts,showpacs,longbibliography, superscriptaddress, 10pt, nofootinbib]{revtex4-2}

\usepackage{graphicx}
\usepackage{verbatim,epsfig,amsmath,amssymb,bm,epsf,psfrag,bbold,amsthm,amsfonts}
\usepackage[toc,page,titletoc]{appendix}
\usepackage[bottom]{footmisc}

\usepackage{subcaption}

\usepackage{enumitem}
\usepackage{soul}
\usepackage{grffile}
\usepackage{framed}
\usepackage{mathrsfs}
\usepackage{esint}
\setlist{nosep}
\usepackage{placeins}
\usepackage[all]{xy}
\usepackage{color}
\usepackage[utf8]{inputenc}
\usepackage{natbib}
\usepackage{tikz-cd}
\usepackage{verbatim}
\usepackage{leftidx}
\usepackage[normalem]{ulem}

\usepackage{notes2bib}

\usepackage{hyperref}
\hypersetup{
    colorlinks=true,
    linkcolor=blue,
    citecolor=blue,
    filecolor=black,
    urlcolor=blue,
}

\usepackage[capitalize]{cleveref} 

\usepackage{xcolor}

\newcommand{\abs}[1]{\left|#1\right|}

\newcommand\beq {\begin{equation}}
	\newcommand\eeq {\end{equation}}
\newcommand\beqa {\begin{equatiobn}\begin{array}}
		\newcommand\eeqa {\end{array}\end{equation}}
\newcommand\bal {\begin{align}}
	\newcommand\eal {\end{align}}
\newcommand{\bea}{\begin{eqnarray}}
	\newcommand{\eea}{\end{eqnarray}}

\theoremstyle{plain}

\theoremstyle{definition}

\theoremstyle{remark}

\usetikzlibrary{arrows,decorations.pathreplacing}
\usetikzlibrary{decorations.pathmorphing}

\tikzset{snake it/.style={decorate, decoration=snake}}\usepackage[T1]{fontenc}

\usetikzlibrary{decorations.markings}

\begin{document}
	
\title{Classical Fractons: Local chaos, global broken ergodicity and an arrow of time}

\author{Aryaman Babbar}
\email{aryaman.babbar@seh.ox.ac.uk}
\affiliation{St Edmund Hall, University of Oxford, Oxford OX1 4AR, United Kingdom}
\author{Ylias Sadki}
\email{ylias.sadki@physics.ox.ac.uk}
\affiliation{Rudolf Peierls Centre for Theoretical Physics, University of Oxford, Oxford OX1 3PU, United Kingdom}
\author{Abhishodh Prakash}
\email{abhishodhprakash@hri.res.in (he/him/his)}
\affiliation{Rudolf Peierls Centre for Theoretical Physics, University of Oxford, Oxford OX1 3PU, United Kingdom}
\affiliation{Harish-Chandra Research Institute, Prayagraj (Allahabad) 211019, India}
\author{S. L. Sondhi}
\email{shivaji.sondhi@physics.ox.ac.uk}
\affiliation{Rudolf Peierls Centre for Theoretical Physics, University of Oxford, Oxford OX1 3PU, United Kingdom}	

\date{\today}

\begin{abstract}

We report new results on classical non-relativistic dipole conserving particles --- fractons. These have been previously shown to  exhibit `Machian' dynamics where the motion of one particle requires the presence of others in its proximity, such that dynamics produces ergodicity breaking steady states characterized by clusters. In this work, we show that although the global state breaks ergodicity, a limited version of ergodic behavior is retained within the clusters which may or may not be chaotic, depending on the nature of the microscopic Hamiltonian. In certain cases, we show that the dynamics can be mapped to that of a billiards particle in various stadiums. We also show that the many-fracton trajectories characteristically exhibit a central time or `Janus point' and thus a generic non-equilibrium bi-directional arrow of time.

\end{abstract}

\maketitle
\tableofcontents

\section{Introduction} 

In this paper, we continue the study of classical particles conserving global multipole moments, begun in \cite{classical_fractons} and continued in \cite{machian_fractons}. 
The original and a continuing motivation for this study is that particles obeying exact multipolar conservation laws are perhaps the simplest examples of ``fractons'', by which we mean particles which exhibit subdimensional mobility—that is, their motion is restricted to manifolds of a dimension that is strictly less than the ambient dimension. 
We hasten to add that there is much larger literature on fractons from various viewpoints covered in several excellent reviews \cite{GromovRadzihovsky2022fractonReview,NandkishoreHermeleFractonsannurev-conmatphys-031218-013604,PretkoChenYou_2020fracton} to which we direct the reader, although familiarity with that literature is not needed to understand our work, which is quite self-contained. 

A second motivation that emerged in the course of the study was that these are naturally nonlinear systems, and they exhibit a surprising number of features contrary to the intuition most physicists develop in engaging with the canonical results of classical mechanics. 
For example, in previous work, we have shown that these fracton systems exhibit attractors and spontaneous translation symmetry breaking in one and two dimensions, which is generally believed to be excluded by the Hohenberg-Mermin-Wagner-Coleman theorem \cite{HohenbergPhysRev.158.383,MerminWagner_PhysRevLett.17.1133,Colemancmp/1103859034}. 
Needless to say, both these properties go along with their lack of ergodicity and, less obviously, with the lack of existence of a proper statistical mechanics—all of the above despite perfectly well-defined Hamiltonians. 
We also provided a heuristic understanding of the structure of our attractors via a generalization of the idea of order-by-disorder, using a measure of non-equilibrium entropy.

In this paper, we continue further along this line of inquiry and examine more carefully the interplay of chaos and emergent integrability in constraining the dynamics of our systems. We find that the late-time attractors generically combine a set of emergent conserved quantities that destroy global ergodicity, and yet are organized in local clusters within which motion is locally ergodic yet not chaotic. This is one of our primary results.

Our second primary result is that the late-time attractors combined with time reversal lead, in our systems, to the phenomenon of Janus points where a generic dynamical trajectory exhibits a minimum in a suitable dynamical variable or non-equilibrium entropy or both. Away from the Janus point, there are two possible ``physical arrows of time'' which are in consonance with or reversed from the time variable that enters the equations of motion. Such Janus points have been introduced into discussions of the fundamental arrow of time in cosmology by \textcite{carroll2004spontaneous} and by Barbour, Koslowski and Mercati \cite{Barbour2014}. This work has been discussed critically and pedagogically in \cite{Goldstein2016} and especially in \cite{Lazarovici_Reichert_arrows_of_time} which makes an important distinction between the entropic/dynamical nature of the two lines of work which is not directly germane to our work. Our Janus points, however, have a strong visual resemblance to those in classical gravitational systems. Indeed, we also refer to a Janus point as a ``big bang'' --- a terminology on which we settled naturally in ignorance of the cosmological literature. 
The cosmological examples are not ergodic systems, and we find it interesting that our very different non-ergodic system also exhibits the phenomenon that generic trajectories have such Janus points. 

Before turning to the details of these results, we first provide a quick recap of the setup and the previous results.

\subsection{Hamiltonians and summary of earlier results}

We begin by introducing classical ``Machian'' Fractons as in \cite{classical_fractons}.
These fractons are Hamiltonian systems which impose a dipole conservation law, leading to fractonic behavior.

We consider $N$ classically identical particles in $d=1$ spatial dimensions with positions $\{{x_i}\}$ and momenta $\{{p_i}\}$ carrying the same $U(1)$ charge; the generalization to higher dimensions is straightforward and we have reported some results on $d=2$ previously \cite{machian_fractons}. 
Translation invariance and dipole moment conservation require that
\begin{equation}
{P} = \sum_{i=1}^N {p_i} \quad \text{and} \quad {D} = \sum_{i=1}^N {x_i}
\end{equation}
have vanishing Poisson brackets with the Hamiltonian. 
More simply, much as the conservation of total momentum $P$ requires our Hamiltonian be independent under uniform spatial translations, $x_i \rightarrow x_i + \phi$, conservation of dipole moment ${D}$ requires invariance under translations of momenta: $p_i \rightarrow p_i + \eta$. 
As a consequence, the Hamiltonian can only depend on position differences and on momentum differences. 
This is a novel feature for fractons and will lead to much of the new physics. 

As usual we try to construct a Hamiltonian by writing down $k$-particle terms with the smallest non-trivial values of $k$. Our symmetries dictate that all lowest order terms must involve two particles:
\begin{equation}
    H = \sum_{i<j} H^p(p_k - p_l) + H^x(x_i - x_j) + H^{xp}(x_i - x_j, p_k - p_l).
\end{equation}
In this general form, the Hamiltonian will be non-local as the first term allows two particles to influence each other at arbitrary distances, which is unphysical. 
To impose locality, we drop the first term and require that the $H^x_{ij}$ and $H^{xp}_{ij}$ fall off with distance: each quadratic term in the Hamiltonian in only ``switched on'' when the two corresponding fractons are sufficiently close to each other. 
Finally, we follow tradition and expand the momentum dependence in a Taylor series about $p_i - p_j = 0$.
The simplest Hamiltonian, quadratic in momenta, that satisfies these conditions is
\begin{equation}
    H = \frac{1}{2} \sum_{i=1}^N \sum_{j=1}^{i-1} \left(p_i - p_j\right)^2 K\left(x_i - x_j\right),
    \label{eqn:main_hamiltonian}
\end{equation}
where $K(x)$, which we will term the \emph{pair inertia function}, imposes locality.
In the following we will take \cref{eqn:main_hamiltonian} to be the Hamiltonian of a system of $N$ dipole conserving fractons, in one spatial dimension. As the inertia of the particles requires them to be close to each other, we refer to this as {\it Machian dynamics} in honor of Mach's principle \cite{Pretko_MachPhysRevD.96.024051}. These fractons were the subject of previous studies \cite{classical_fractons, machian_fractons}, and we summarize their main features here.
\medskip
\begin{enumerate}
\itemsep.5em
\item First, these fractons appear \emph{dissipative} \cite{classical_fractons}, in an apparent contradiction of Liouville's theorem, which forbids attractors in Hamiltonian systems.
This is clearly seen for systems of two fractons, which generically separate out to a fixed distance, and motion comes to a halt, reminiscent of a system with friction.
Of course, as a rigorous theorem, Liouville's theorem cannot be violated: the apparent contradiction is resolved by understanding the attractor is only present in position-velocity space.

\item Conventional Hamiltonian systems have a linear relationship between position and velocity for each particle.
However, velocities of individual fractons involve the pair inertia function, and momenta of all other fractons.
As the apparent attractor is approached, shrinking in position space, momenta of particles diverge, conserving true phase space volume, in compliance with Liouville's theorem.

\item 
Ergodicity is found to be broken in an unusual manner \cite{machian_fractons}.
Late time states always converge to attractors, with the emergence of new conserved quantities.
Most interestingly, late time states generically lead to the breaking of translation symmetry into crystalline states, even in low dimensions, where the naive invocation of the Hohenberg-Mermin-Wagner-Coleman theorem would forbid such a breaking of a continuous symmetry.

\end{enumerate}

\subsection{Preview of coming attractions}

In Section II, we provide a complete characterization of the three-fracton problem. 
We show how trajectories initiated in different regions of phase space evolve and demonstrate the emergence of a central time or ``Janus point'' that hints at a fundamental arrow of time.

Section III examines the four-fracton system, where we illustrate the interplay between local chaos and global broken ergodicity. 
We show late-time clustering states reduce to billiard-like motion in confined regions of phase space, while conserving certain quantities that prevent full ergodicity. 
For compact pair inertia functions $K$, we demonstrate that these states exhibit regular, integrable dynamics, while systems with non-compact $K$ can display chaotic behavior.

In Section IV, we explore the emergence of an arrow of time. 
We introduce a complexity measure that increases monotonically away from a central Janus point, analogous to the behavior seen in gravitational systems. 
Despite the time-reversal symmetry of the underlying dynamics, we demonstrate how this measure provides a natural direction for time.

\section{Three fractons}
\subsection{Review of formulation and known results}
Let us begin with a review of the general formulation for the Hamiltonian dynamics of $N$ particles.
Although we have $2N$ position and momentum degrees of freedom, the Hamiltonian is independent of both total momenta and total position (dipole moment). Correspondingly, we write the Hamiltonian in terms of the `reduced coordinates', introduced in \cite{classical_fractons} as follows:
\begin{align}
        q_m =
\begin{cases} 
    \frac{\sum_{j=1}^m x_j - m x_{m+1}}{\sqrt{m(m+1)}} & \text{if } 0 < m < N-1 \\
    \frac{\sum_{j=1}^N x_j}{\sqrt{N}} & \text{if } m = N.
\end{cases} \label{eqn:reduced_q} \\ 
        \pi_m =
\begin{cases} 
    \frac{\sum_{j=1}^m p_j - m p_{m+1}}{\sqrt{m(m+1)}} & \text{if } 0 < m < N-1 \\
    \frac{\sum_{j=1}^N p_j}{\sqrt{N}} & \text{if } m = N.
\end{cases} \label{eqn:reduced_pi}
\end{align}

The Hamiltonian does not depend on $q_N$ and $\pi_N$, which are the conserved quantities, so we can now express the Hamiltonian in terms of only $2N-2$ coordinates. In particular, for $N=3$, we can express the Hamiltonian in terms of $q_1$, $q_2$, $\pi_1$ and $\pi_2$ only:

\begin{multline}
    H = \pi_1^2 K(\sqrt{2}q_1) + 
    \frac{(\sqrt{3}\pi_2 - \pi_1)^2}{4} K\left( \frac{\sqrt{3} q_2 - q_1}{\sqrt{2}}\right) + \\ 
    \frac{(\sqrt{3}\pi_2 + \pi_1)^2}{4} K\left( \frac{\sqrt{3} q_2 + q_1}{\sqrt{2}}\right).
\end{multline}

Let us summarize aspects of the three-particle dynamics presented in Ref.~\cite{classical_fractons}. The results we present next are valid for compact pair inertia functions: that is, $K(x) = 0$ strictly, for  $\abs{x} \geq a$ for some $a$. For our arguments, it is sufficient to consider a box pair inertia function, i.e. 

\begin{equation}
        K(x) = 
\begin{cases} 
    1 & \text{if } |x| < 1 \\
    0 & \text{if } |x| \geq 1.
\end{cases}
\end{equation}

To characterize the trajectories, we must first classify the different configurations of fractons. As shown in Ref.~\cite{classical_fractons}, we can divide the generalized position $(q_1, q_2)$ plane into regions where a varying number of pair-inertia terms are ``switched on'' (i.e., equal to $1$).
The regions are shown in \cref{fig:3fractrajectory}.
Region $3$ is hexagonal, where all three pair inertia functions are switched on. 
Regions $2_{a-f}$ have two $K$s switched on, and consist of six equilateral triangles. 
Regions $1_{a-f}$ have exactly one $K$ switched on, and similarly regions $0_{a-f}$ have zero $K$s switched on.
We can further divide regions $1_{a-f}$ into regions accessible (shaded) or inaccessible (unshaded) to trajectories starting in Region 3.

As shown in Ref.~\cite{classical_fractons}, analyzing trajectories of motion is an exercise in reflection and transmission rules at the boundaries between regions.
Trivially following from the equations of motion, away from boundaries, trajectories in the $(q_1, q_2)$ plane travel in straight lines.
However, upon encountering a boundary, trajectories jump discontinuously in momenta. 
We calculate the momenta after these jumps by noticing that at each boundary there exists a linear combination of the reduced momenta that is conserved before and after the jump. 
These equations, detailed in \cref{appendix:three_fracton}, in addition to the conservation of energy, give us enough information to get a quadratic equation governing the momenta after the jump. 
A reflection, as opposed to transmission into the other region, will occur if there are no real solutions to the system of equations.

\subsection{Complete characterization}
\begin{figure*}[htbp]
    \centering
    \includegraphics[width=\textwidth]{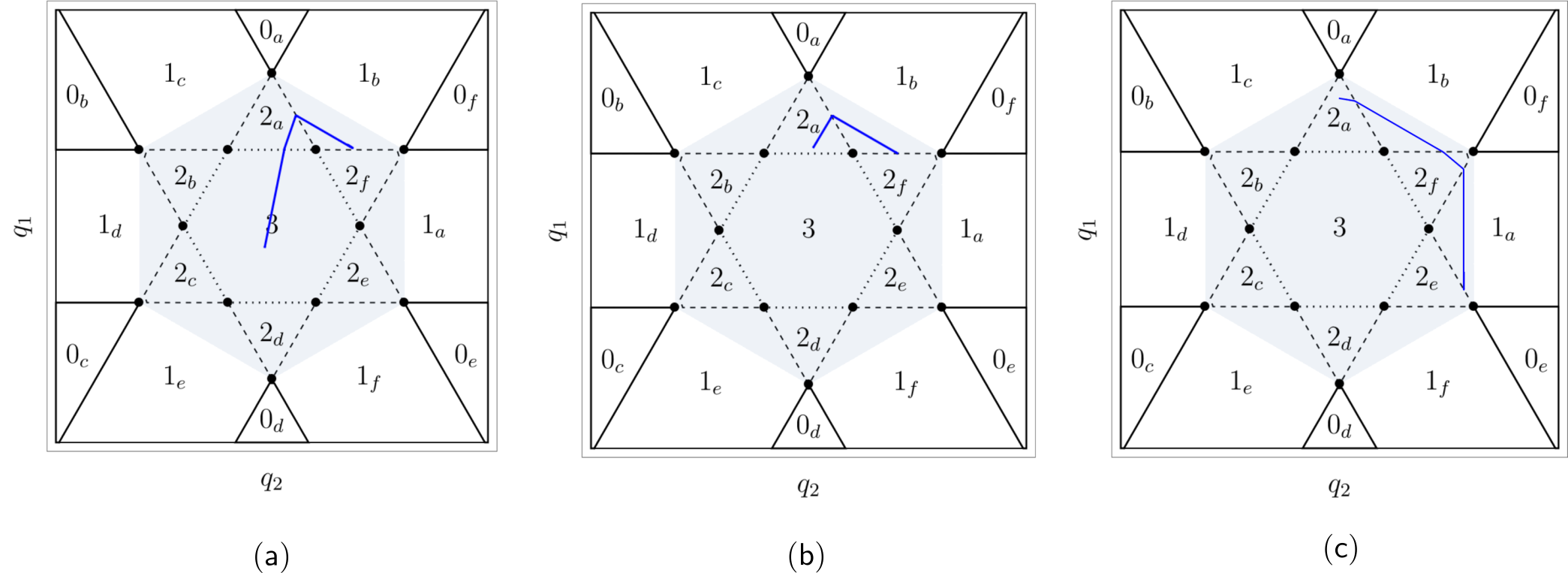}
    \caption{ Three-fracton trajectories. $3$, $2_{a-f}$, $1_{a-f}$ and $0_{a-f}$ represent regions in the reduced coordinate space where 3, 2, 1 and no pairs of particles are within Machian range respectively. A generic trajectory starting in $3$ will always enter a unique $2$ region and end up in a $1$ region after exactly two `refractions' as shown in (a). A trajectory starting in one of the $2$ regions can end in the first $1$ region it encounters through a single refraction (b) or or re-enters a $2$ region and gets trapped in the next $1$ region, after two refractions (c). }
    \label{fig:3fractrajectory}
\end{figure*}

We now fully characterize three-particle trajectories based on the initial positions and momenta.
Given the permutation symmetry of the system, the following holds in generality for any of the sub-regions $2_{a-f}$.
We enumerate the following possible starting conditions:

\begin{enumerate}
\itemsep0.5em
\item (\cref{fig:3fractrajectory}a) A trajectory starting in Region $3$, moving towards Region $2_a$, will always pass into $2_a$. 
Then it will jump from $2_a$ to the first sub-region of $1$ it encounters, and remain trapped there.

\item (\cref{fig:3fractrajectory}b, c) A trajectory starting in Region $2_a$, moving towards Region $1_b$, either remains trapped in $1_b$ (depending on the initial momenta), or it escapes to $2_f$, from where it necessarily goes into $1_a$, and is trapped there. 

\item A trajectory starting in Region $1_a$ either gets trapped in $1_a$ or will reflect off the $1_a$-$2_e$ and $1_a$-$2_f$ boundaries a finite number of times, before escaping to Region $2_e$ or $2_f$. 
The number of such reflections depends on the ratio of reduced momenta $\pi_2/\pi_1$ at the start of the trajectory. 
Once the trajectory has escaped to $2_e$ or $2_f$, it ends up trapped in one of the sub-regions of $1$ as described. 

\end{enumerate}

The analysis of trajectories in reduced coordinates provides a powerful tool to characterize trajectories for few body systems. 
Indeed, in certain late time states, the trajectories act as billiards in the reduced space (\cref{sec:four_fractons}).
Interestingly, time reversing a Type 1 trajectory (i.e.\ one that starts in Region 3) will produce a trajectory that starts in Region $1$, passes through Region $3$, then settles into a different sub-region of $1$.
Passing through the central Region $3$ hints towards a central time (``Janus point'' \cite{Barbour2014}) where particles are maximally close to each other.
In \cref{sec:arrow_of_time} we show this intuition generalizes to the $N$ body problem, generating an emergent arrow of time.

\section{Four Fractons} \label{sec:four_fractons}

As a many body interacting system, fractons are at odds with conventional expectations of chaotic trajectories and ergodicity.
We seek to characterize whether trajectories of classical Machian fractons are chaotic, and the implications on ergodicity.
Properties of chaos and ergodicity are sensitive to the form of the pair inertia function $K$.
We consider three cases:
\smallskip
\begin{enumerate}
\itemsep0.5em
    \item Box $K$: $K(x) = 1$ for $\abs{x} < 1$, and $0$ for $\abs{x} \geq 1$.
    \item Compact $K$: $K(x) = 0$ for $\abs{x} \geq 1$, but generally may take any value for $\abs{x} < 1$.
    \item Non-compact K: $K(x) \rightarrow 0$ only at $x \rightarrow \pm \infty$.
\end{enumerate}

\smallskip

We first focus on the box $K$, where we rigorously understand dynamics by analyzing trajectories in reduced coordinates.
As we shall show, most results from the box $K$ case carry over to the general compact $K$ case.
We then analyze non-compact $K$, which exhibit chaos.

For generic initial conditions in the ``big-bang'' initial state (i.e. all four particles with all pair inertia terms $K$ ``switched on''), late time states involve four clusters, three clusters, or two clusters, but never one.

Two cluster states are the most interesting, being either the 3--1 state or the 2--2 state.
We denote as 3--1 the state with 3 particles in the first cluster, and 1 in the second cluster.
The first three particles undergo motion within the cluster, even at late times, whereas the fourth particle is frozen in position. 
Similarly, we denote 2--2 as the state with two clusters, containing two fractons each.
All particles undergo motion at late times.

For completeness, we now briefly mention the three cluster and four cluster late time states.
The four cluster case simply has all fractons at rest.
The three cluster case, also called the 1--2--1 state, corresponds to two fractons at rest, with the other two moving within one cluster.
This case is similar to motion in any of the sub-regions of Region 1 of the three fracton problem -- the difference being that the boundaries of reflection change. 

We now turn to the most interesting four fractons cases: two clusters, 3--1 or 2--2.
We find motion is in fact \emph{not} chaotic within each cluster, though remaining ergodic in a restricted region of phase space.
In \cref{sec:lyapunov_chaos} we consider general pair inertia functions $K$ with infinite ranged support, where motion will become chaotic.

\subsection{Four fracton reduced coordinates}

We use the reduced coordinates of Equations \eqref{eqn:reduced_q} and \eqref{eqn:reduced_pi} to reformulate the four fracton Hamiltonian.

\begin{align}
    q_1 &= \frac{x_1 - x_2}{\sqrt{2}},  \nonumber  \\
    q_2 &= \frac{x_1 + x_2 - 2 x_3}{\sqrt{6}}, \nonumber \\
    q_3 &= \frac{x_1 + x_2 + x_3 - 3 x_4}{\sqrt{12}}, 
\end{align}
and similarly for the reduced momenta $\pi_i$ in terms of the momenta $p_j$. 
In these coordinates, the Hamiltonian becomes:

\begin{widetext}
\begin{align} \label{eq:ham}
&H = H(x_1, x_2, x_3, x_4; p_1, p_2, p_3, p_4) \rightarrow H(q_1, q_2, q_3; \pi_1, \pi_2, \pi_3) \nonumber\\ &= \pi_1^2 K\left(\sqrt{2} q_1\right) +
\frac{1}{4}\left(\sqrt{3} \pi_2 + \pi_1\right)^2 K\left(\frac{\sqrt{3} q_2 + q_1}{\sqrt{2}}\right) + 
\frac{1}{4}\left(\sqrt{3} \pi_2 - \pi_1\right)^2 K\left(\frac{\sqrt{3} q_2 - q_1}{\sqrt{2}}\right) \nonumber \\
&+ \frac{1}{18} \left(\sqrt{12} \pi_3 - \sqrt{6}\pi_2\right)^2 K\left(\frac{\sqrt{12} q_3 - \sqrt{6} q_2}{3}\right) + 
\frac{1}{72}\left(3 \sqrt{2} \pi_1 + \sqrt{6}\pi_2 + 2 \sqrt{12} \pi_3\right)^2 K\left(\frac{3 \sqrt{2} q_1 + \sqrt{6}q_2 + 2 \sqrt{12} q_3}{6}\right) \nonumber \\
&+ 
\frac{1}{72} \left(-3 \sqrt{2} \pi_1 + \sqrt{6}\pi_2 + 2 \sqrt{12} \pi_3\right)^2 K\left(\frac{-3 \sqrt{2} q_1 + \sqrt{6}q_2 + 2 \sqrt{12} q_3}{6}\right),
\end{align}
\end{widetext}
which explicitly involves six $K$ terms, corresponding to the six possible pairings of four fractons.

\subsection{3--1 late time state} \label{sec:31state}

We now find an easy-to-visualize 3--1 late time state in reduced position phase space, and then the results will apply to all other 3--1 states by symmetry. 
As for three fractons, we now explicitly use box $K$.

Without loss of generality, we choose the fourth particle's position $x_4$ to be in the isolated cluster, with $x_1$, $x_2$, $x_3$ in the other cluster. 
Further, we choose $x_1 - x_4, \, x_2 - x_4, \, x_3 - x_4 \geq 1$. 
If we now visualize this in reduced coordinate space, we are restricted to the intersection of the region defined by $\sqrt{12} q_3 - \sqrt{6} q_2 \geq 3$, $3 \sqrt{2} q_1 + \sqrt{6} q_2 + 2 \sqrt{12} q_3 \geq 6$ and $-3 \sqrt{2} q_1 + \sqrt{6} q_2 + 2 \sqrt{12} q_3 \geq 6$, and $|x_i - x_j| \leq 1$, where $i$ and $j$ run from 1 to 3. \\

In the Hamiltonian (\cref{eq:ham}), only the first three terms are non-zero, so the equations of motion are

\begin{align} \label{eq:eomqtriangle}
    \Dot{q_1} &= 3 \pi_1, \nonumber \\
    \Dot{q_2} &= 3 \pi_2, \nonumber \\
    \Dot{q_3} &= 0.
\end{align}

\begin{figure*}[t]
    \centering
    \begin{subfigure}[b]{0.3\textwidth} 
        \centering
        \includegraphics[width=\textwidth]{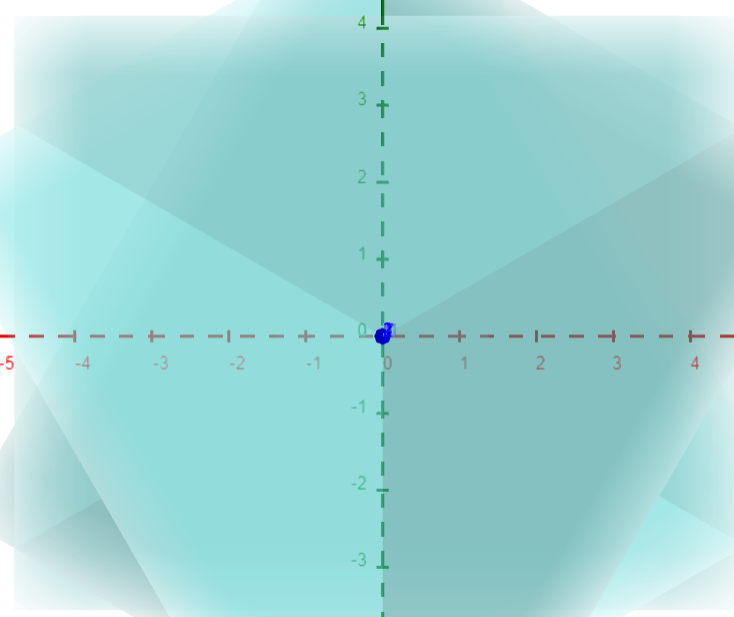}
        \caption{}
        \label{fig:top_view_planes}
    \end{subfigure}
    \hfill
    \begin{subfigure}[b]{0.3\textwidth} 
        \centering
        \includegraphics[width=\textwidth]{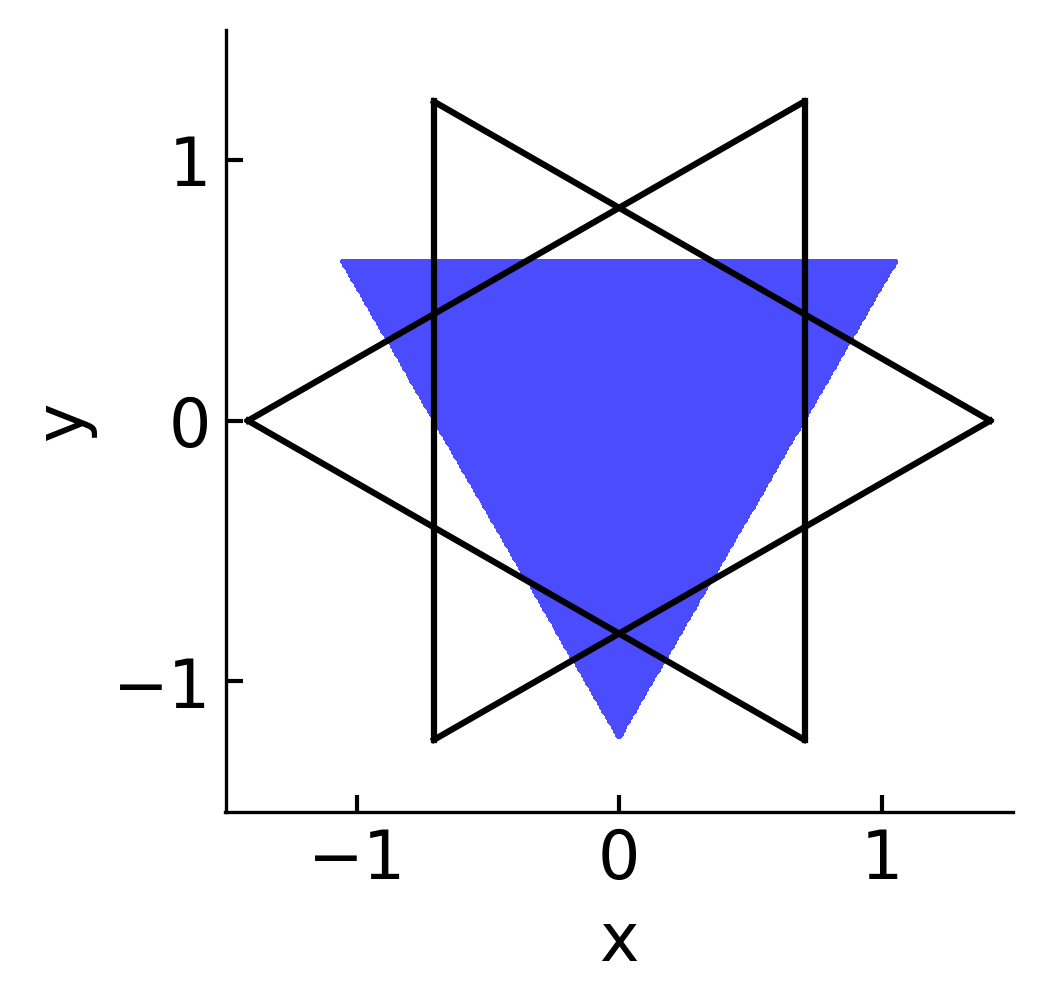}
        \caption{}
        \label{fig:large_tri_hex}
    \end{subfigure}
    \hfill
    \begin{subfigure}[b]{0.3\textwidth} 
        \centering
        \includegraphics[width=\textwidth]{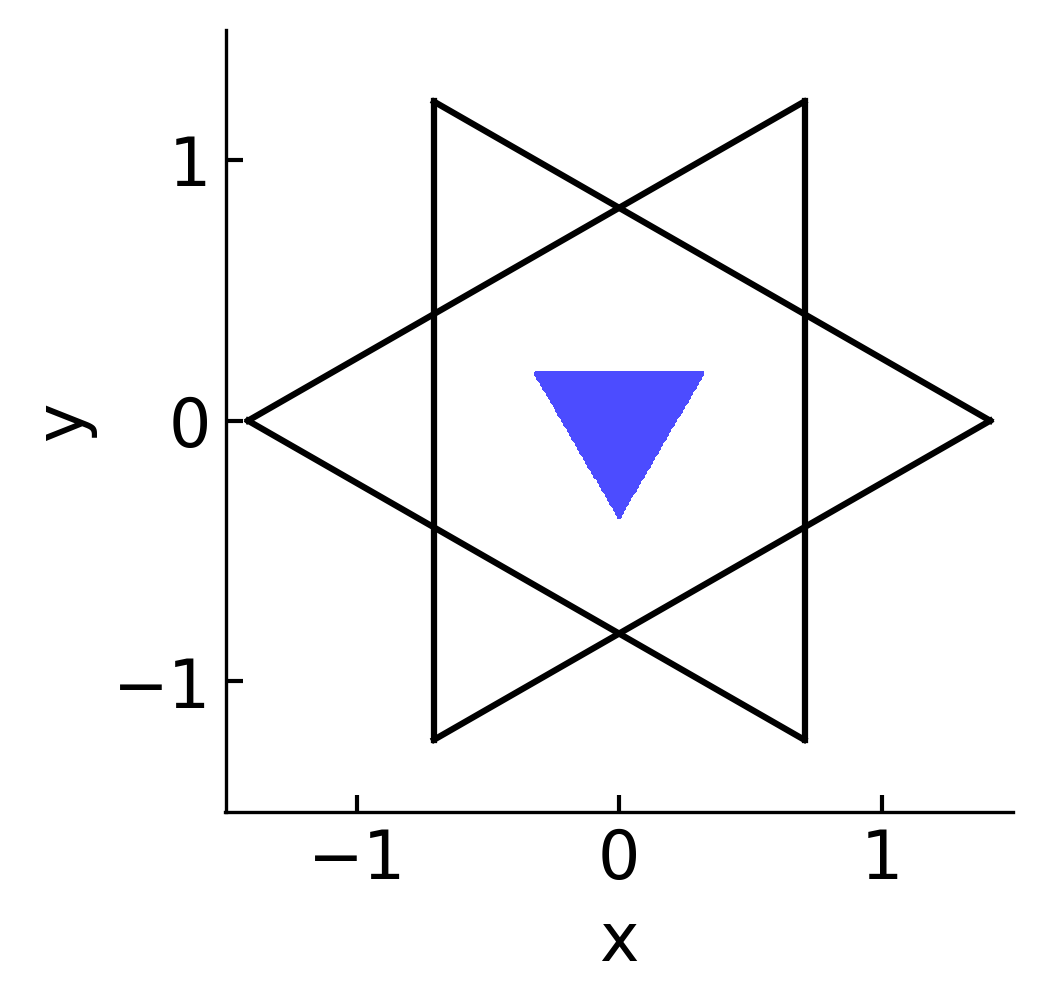}
        \caption{}
        \label{fig:small_tri_hex}
    \end{subfigure}

    \caption{Visualizing the 3--1 clustering in the late-state dynamics for four fractons discussed in \cref{sec:31state}: The late-time steady state dynamics is confined to a plane in the reduced coordinates $q_1, q_2, q_3$ which we denote as the $x-y$ plane. $z$ points out of the page. The dependence of $x,y,z$ on $q_1, q_2, q_3$ depends on the precise configuration of clustering that has set in. (a) Three planes that define an infinitely based pyramid, with the open base at $z = \infty$. The volume is defined by $x_1 - x_4, \, x_2 - x_4, \, x_3 - x_4 \geq 1$, which is a specific case where the 3--1 state has the 4th particle isolated. Any late time state is then defined by a two-dimensional slice (equilateral triangle), $z = \textrm{const}$ within this volume.
    (b),(c): Various configurations of the equilateral triangle, which bounds trajectories for 3--1 late time states.}
    \label{fig:4_1}
\end{figure*}

Hence, the trajectory is restricted to being parallel to the $q_1 - q_2$ plane. 
Therefore, we consider slices parallel to the $q_1 - q_2$ plane.
In the constrained region, these planes become equilateral triangles, whose size characterizes the dynamics (\cref{fig:top_view_planes} is a view of the region from the positive z-direction).

We now recycle the $q_1$, $q_2$ visualization for the three fracton problem, and apply it to the three fractons in the main cluster.
Overlaying the equilateral triangle that bounds the four fracton dynamics, we see (\cref{fig:large_tri_hex,fig:small_tri_hex}) the triangle is either \emph{large}, so it overlaps outside the central region, or \emph{small}, so it is contained entirely within the central region.

For the large triangle case (\cref{fig:large_tri_hex}), the triangular cross-section includes the sub-regions where we get three Machian clusters (Regions $1_a$ to $1_f$), so such a trajectory would pass through states with \emph{two} particles at rest.
This is never observed in simulations of 3--1 states. 
Hence, the late time 3--1 state, for box $K$, is confined by the small triangle: \cref{fig:small_tri_hex}.

For general compact $K$, the same late time states are observed: 3--1 clustering always involves the small triangle state.

\begin{figure}
\centering
\begin{subfigure}{8.6cm}
  \centering
  \caption{}
  \includegraphics[width=8.6cm]{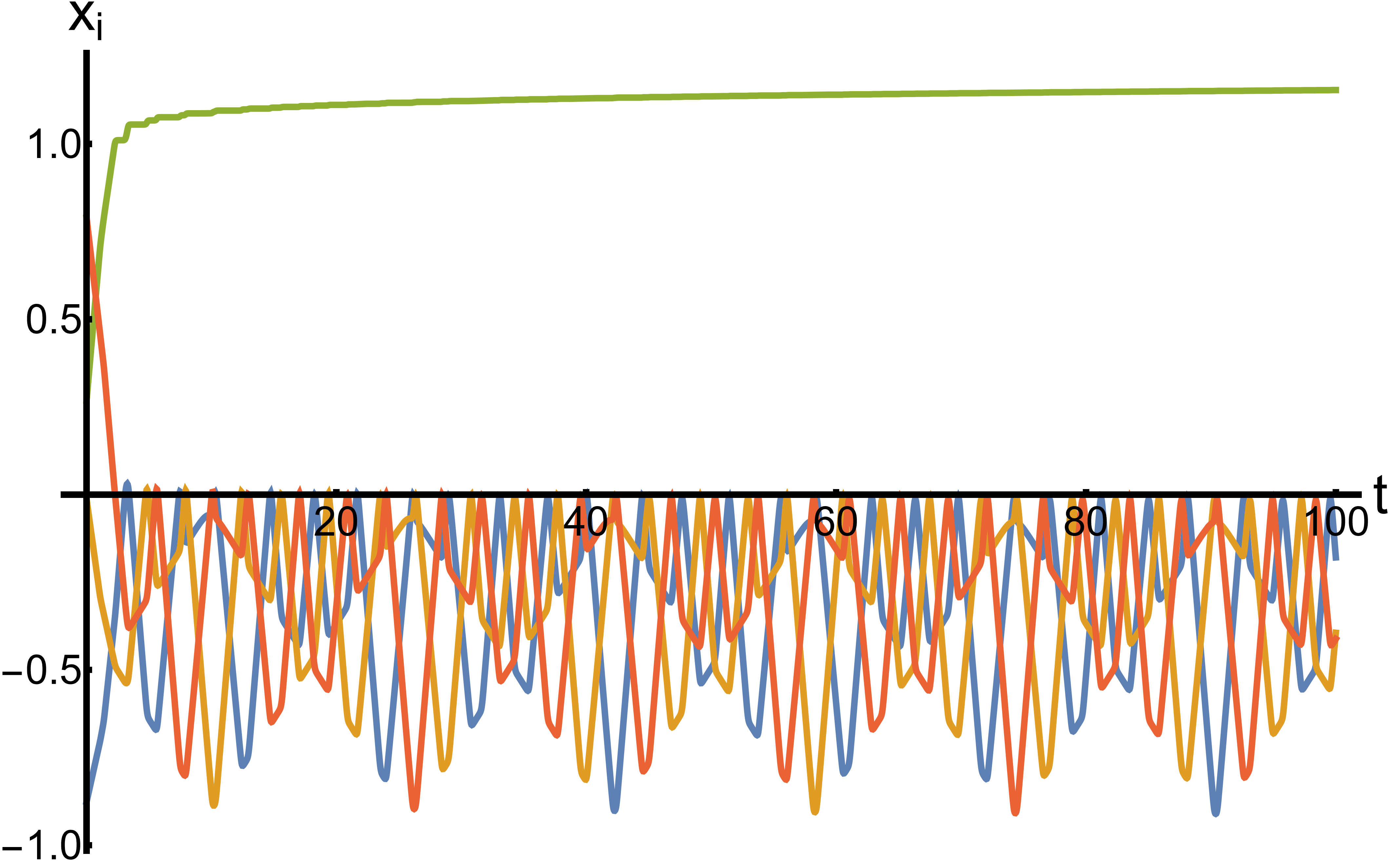}
  \label{fig:31pos}
\end{subfigure}
\hfill
\begin{subfigure}{8.6cm}
  \centering
  \caption{}
  \includegraphics[width=8.6cm]{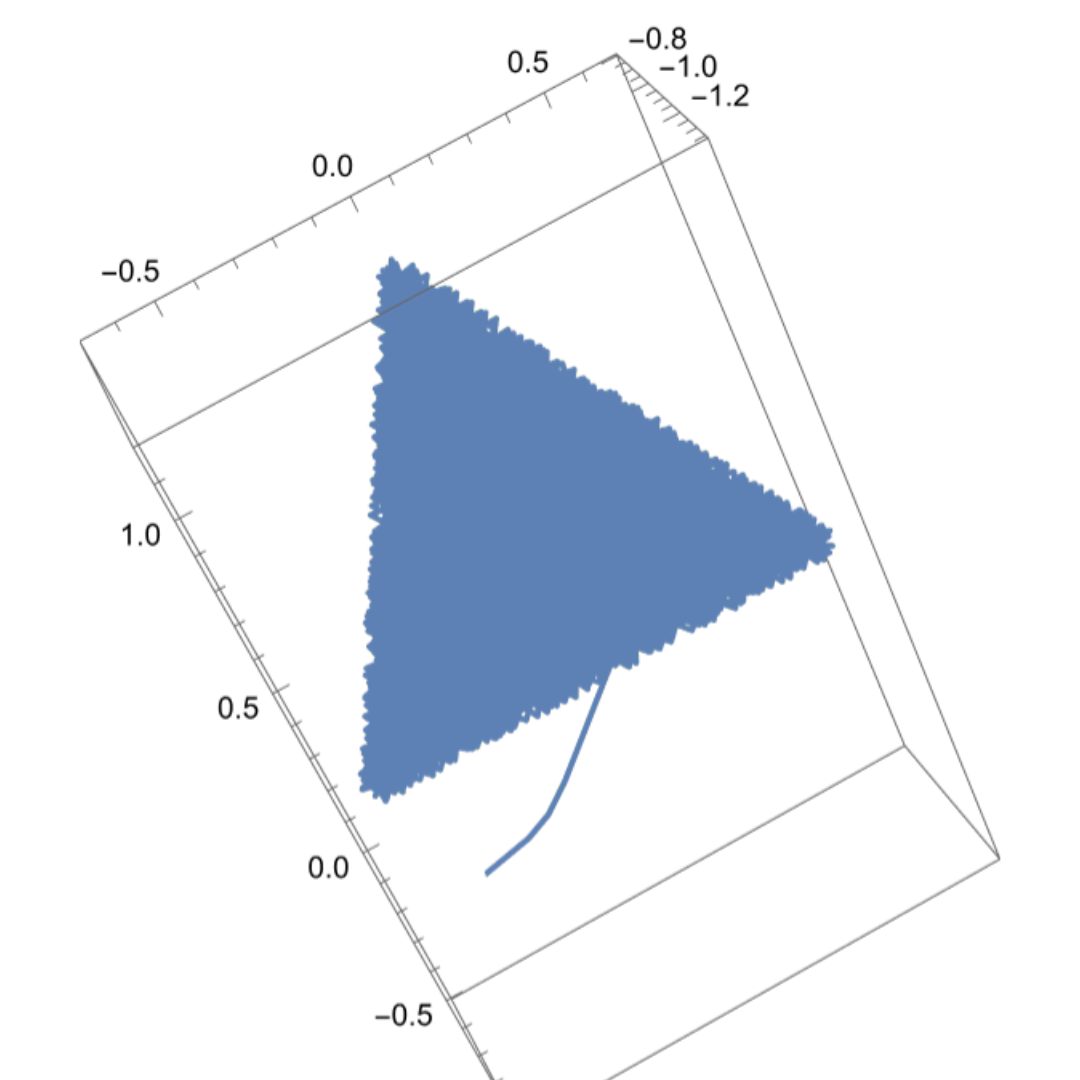}
  \label{fig:31sq}
\end{subfigure}
\caption{(a): Time-evolution of positions of each particle in a 3--1 trajectory discussed in \cref{sec:31state}. (b): the trajectory in reduced position coordinates. }
\label{fig:3_1}
\end{figure}

We have reduced the 3--1 four body problem to a constrained problem on a triangle.
Now we ask, how does the trajectory behave within this triangle?
For box $K$, trajectories will undergo motion in a straight line until they strike an edge of the triangle. 
Consider such an event: assume it hits the edge parallel to the x-axis, i.e.\ $\sqrt{12} q_3 - \sqrt{6} q_2 = 3$. 
Assuming the trajectory does not escape the triangle, we have three constants of motion.
First, the Hamiltonian $H_{3-1} = 3/2 (\pi_1^2 + \pi_2^2) = \mathrm{const}$.
Generalizing the approach in \cite{classical_fractons}, we derive from the full Hamiltonian \cref{eq:ham} at the boundary: $\pi_1 = \textrm{const}$ and $\sqrt{2} \pi_2 + \pi_3 = \textrm{const}$.
Writing the new $\pi_i^\prime$ in terms of the old $\pi_i$, we have: $\pi_1' = \pi_1$, $\pi_2' = -\pi_2$ and $\pi_3' = 2 \sqrt{2} \pi_2 + \pi_3$. 
The component of the velocity parallel to the edge remains constant, while the component perpendicular to the edge reverses sign --- this is characteristic of a dynamical billiard. 
Strikingly, the four body interacting system reduces to billiards exploring a triangle.

While $\pi_1$ and $\pi_2$ behave like the components of momentum of a dynamical billiard in an equilateral triangular domain, $\pi_3$ does not. 
$\pi_3$ changes only when the trajectory strikes the boundary of the triangle, and the jump in $\pi_3$ can be computed as detailed above. 
This can be generalized to any $3-1$ late time state.

\subsection{2--2 late time state} \label{sec:22state}

The two cluster late time states for the four fracton problem are distinguished into the 3--1 and 2--2 cases.
We have shown the motion in the 3--1 case reduces to billiard motion within a triangle.
We now perform a similar analysis for the 2--2 case.
Without loss of generality, we choose the first two particles to be in the first cluster, and the last two particles in the second cluster.
The Hamiltonian is then given by

\begin{equation}
    H_{2-2} = \pi_1^2 + \frac{\left(\sqrt{2} \pi_3 - \pi_2\right)^2}{3},
\end{equation}

so the equations of motion are 

\begin{align}
    \Dot{q_1} &= 2 \pi_1, \nonumber \\
    \Dot{q_2} &= -\frac{2}{3} (\sqrt{2} \pi_3 - \pi_2), \nonumber \\
    \Dot{q_3} &= \frac{2 \sqrt{2}}{3} (\sqrt{2} \pi_3 - \pi_2).
\end{align}

\begin{figure}
    \centering
    \includegraphics[width=8.6cm]{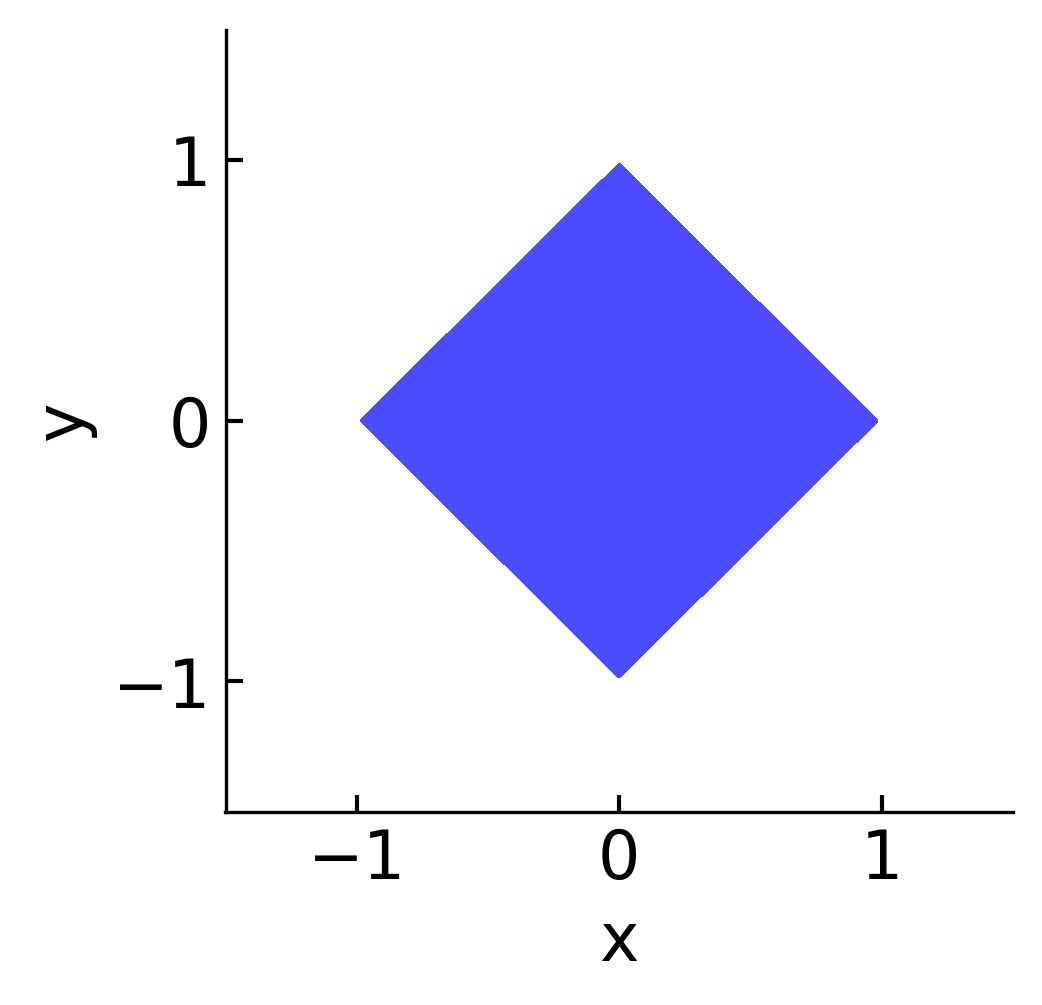}
    \caption{The intersection of regions $x_1 - x_3 \geq 1, x_2 - x_3 \geq 1, x_2 - x_4 \geq 1$ and $x_1 - x_4 \geq 1$ in $x-y$ coordinates, defined in \cref{eq:x_y_definition}. The 2--2 late time states discussed in \cref{sec:22state} have trajectories confined within this region.}
    \label{fig:4_2}
\end{figure}

This clustering configuration demands $-\sqrt{1} \leq \sqrt{2} q_1 \leq 1$ and $-3 \leq \sqrt{12} q_3 - \sqrt{6} q_2 \leq 3$. 
We also choose $x_1 - x_4, \, x_2 - x_4, \, x_1 - x_3, \, x_2 - x_3 \, \geq 1$. 
The equations of motion imply the trajectory moves on the plane $\sqrt{2} q_2 + q_3 = \text{const}$, so our motion is once again on a plane. 
We now change coordinates: we set 
\begin{align}
\label{eq:x_y_definition}
x &= q_1, \nonumber \\
y &= (\sqrt{2} q_3 - q_2)/\sqrt{3}, \nonumber \\
C &= (\sqrt{2} q_2 + q_3)/\sqrt{3},
\end{align}
where $C$ is constant.
In the $x - y$ plane, the region $x_1 - x_4, \, x_2 - x_4, \, x_1 - x_3, \, x_2 - x_3 \, \geq 1$ looks like a square, as can be seen in \cref{fig:4_2}. 
In terms of these new coordinates, we have the following equations of motion

\begin{figure}
\centering
\begin{subfigure}{8.6cm}
  \centering
  \caption{}
  \includegraphics[width=8.6cm]{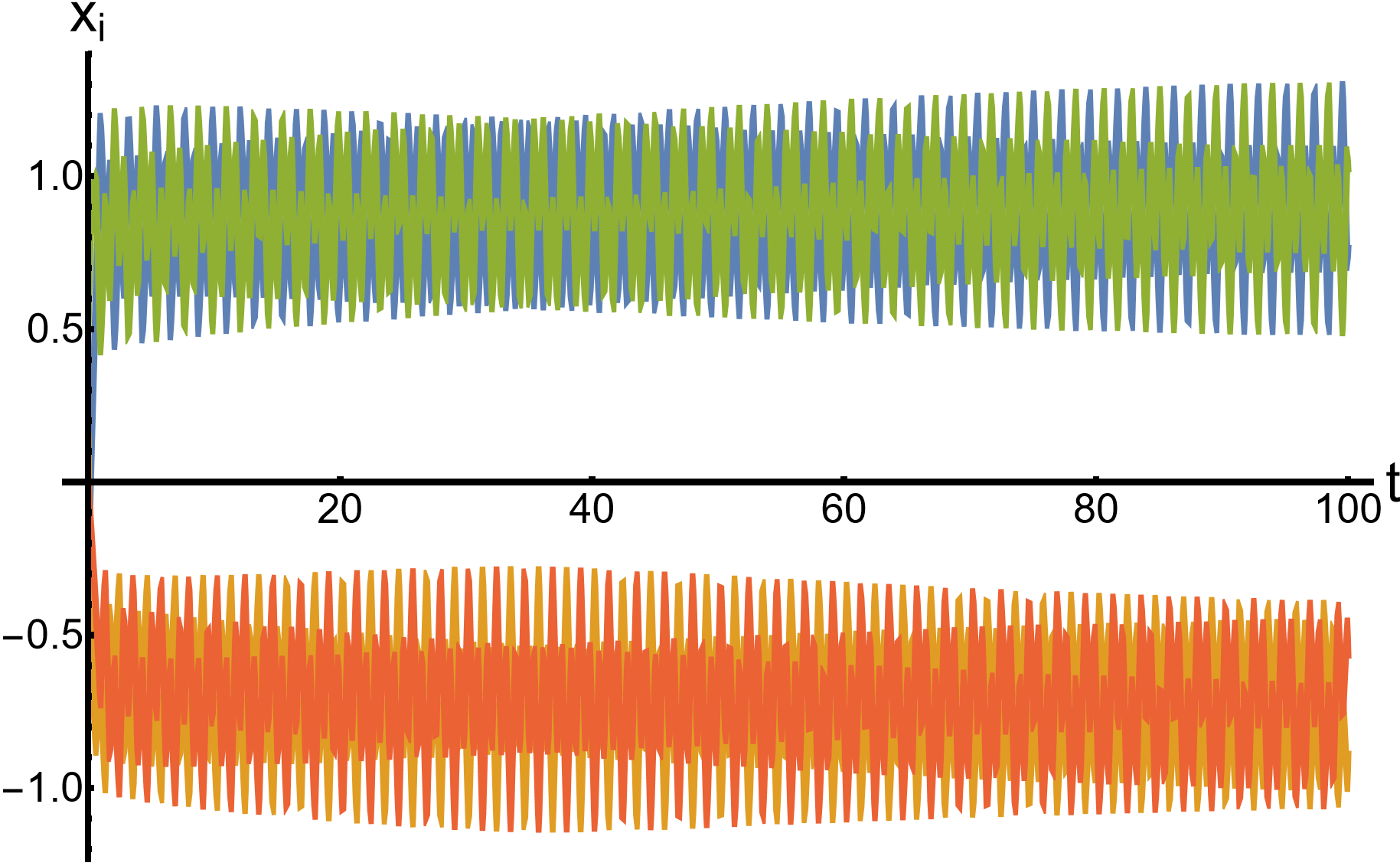}
  \label{fig:22pos}
\end{subfigure}
\hfill
\begin{subfigure}{8.6cm}
  \centering
  \caption{}
  \includegraphics[width=8.6cm]{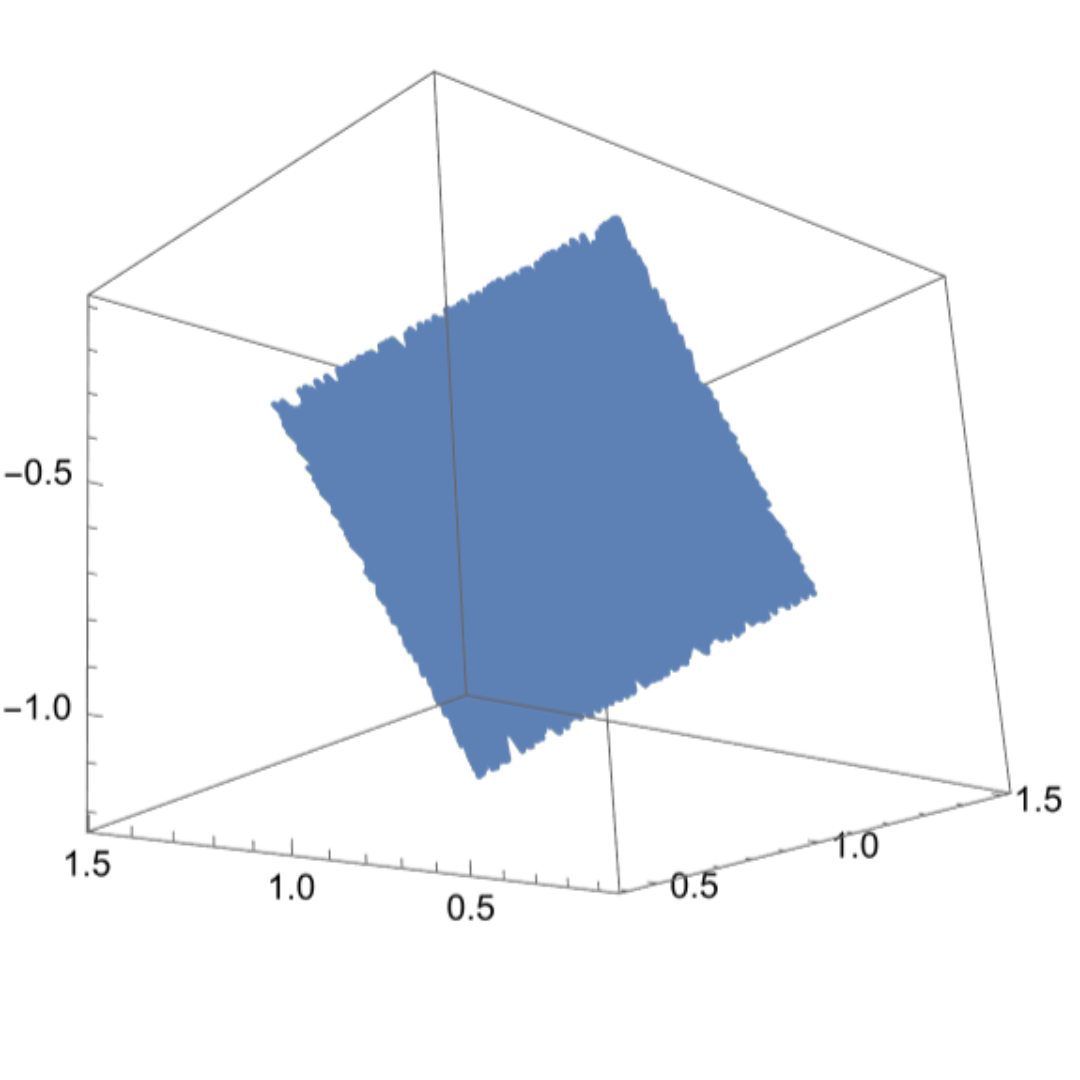}
  \label{fig:22sq}
\end{subfigure}
\caption{(a): Time evolution of particle positions in a 2--2 trajectory discussed in \cref{sec:22state} (b): the trajectory in reduced position coordinates. }
\label{fig:2_2}
\end{figure}

\begin{align}
    \Dot{x} &= 2 \pi_1, \nonumber \\
    \Dot{y} &= 2 \frac{\sqrt{2} \pi_3 - \pi_2}{\sqrt{3}}.
\end{align}

Consider the trajectory striking one of the walls of this square domain. 
The top-left edge in \cref{fig:4_2} corresponds to the wall formed by $\sqrt{3} q_2 + q_1 = \sqrt{2}$. 
Similarly to the 3--1 case, the Hamiltonian $H_{2-2} = \pi_1^2 + {\left(\sqrt{2} \pi_3 - \pi_2\right)^2}/{3} = \mathrm{const}$.
The full Hamiltonian \cref{eq:ham} yields: $\pi_3 = \mathrm{const}$ and $\sqrt{3} \pi_1 - \pi_2 = \mathrm{const}$.
Solving for the new coordinates $\pi_i'$ in terms of the old ones $\pi_i$:
\begin{align}
    \pi_1' &= \frac{\sqrt{2} \pi_3 - \pi_2}{\sqrt{3}}, \nonumber \\
    \pi_2' &= \sqrt{2} \pi_3 - \sqrt{3} \pi_1, \nonumber \\
    \pi_3' &= \pi_3.
\end{align}

Fascinatingly, we have that $\Dot{x'} = \Dot{y}$ and $\Dot{y'} = \Dot{x}$, so in the new $x,y$ phase space coordinates, the trajectory perfectly reflects off the wall. 
Similarly to the $3-1$ case, we can construct three linearly independent combinations of the three momenta, two of which ($x$ and $y$) behave like the momentum components of a billiard, and one of which does not: it only changes when the trajectory strikes the boundary of the square.
Once again, the trajectory behaves like a dynamical billiard, only this time in a square domain. 

\section{Emergent billiards and local chaos}

\subsection{Billiard in a polygonal domain} \label{sec:billiards}

We have shown late time 3--1 and 2--2 states for the four fracton problem reduce to billiards.
Billiards are a well studied problem: we now connect our system to known results of billiard systems \cite{Naydenov2013, ledberg2018introduction}.
Notably, Lyapunov chaos, being exponential divergence of trajectories, is \emph{not} observed in polygonal billiards \cite{Naydenov2013}.
Hence, intuitively we would expect fractonic billiards to exhibit no chaos.
We now prove this.

The integrals of motion constrain the dynamics. 
We now enumerate all of them.
After multiple reflections of billiards in a polygon with angles being rational multiples of $\pi$, the set $\Gamma$ of the number of possible (orthogonal) transformations of the velocity we may have during any trajectory is finite \cite{ledberg2018introduction} (for irrational polygons, $\Gamma$ is an infinite set).
Hence, there is a finite number of directions that the billiard can move in --- this forms an integral of motion. 
In total, there are three integrals of motion:

\smallskip

\begin{enumerate}
\itemsep0.5em
    \item The plane containing the trajectory: $q_3 = \textrm{const}$ for 3--1 states and $\sqrt{2} q_2 + q_3 = \textrm{const}$ for 2--2 states.
    \item The magnitude of velocity: $H_{3-1} = 3/2 (\pi_1^2 + \pi_2^2) = \mathrm{const}$ for 3--1 states. $H_{2-2} = \frac{1}{4} (\dot{x}^2 + \dot{y}^2) = \textrm{const}$ for 2--2 states.
    \item The finite number of directions of the billiard. In an extended zone scheme, with the squares or triangles being tiled, this maps onto a fixed velocity direction.
\end{enumerate}
Hence our system with six degrees of freedom has three integrals of motion.
A well known fact states that a Hamiltonian system with a $2N$ dimensional phase, and $N$ integrals of motion, is \emph{Liouville integrable} \cite{Arutyunov2019}.
So the 3--1 cluster (billiard in an equilateral triangle) and the 2--2 cluster (billiard in a square), have regular and integrable dynamics --- again, no chaos!

For $N>4$ particles, at late times, particles remain within clusters, with close position and momentum values.
This hints towards clustered states being robust to perturbations in phase space: we do not expect perturbations to grow exponentially in time.

\subsection{Lyapunov chaos in the four fracton problem with unbounded $K$} \label{sec:lyapunov_chaos}

The four fracton problem maps on to billiards, for compact box pair inertia $K$, exhibiting regular, integrable dynamics with no chaos. 
We now consider pair inertia functions with infinite range (``tails''), which do exhibit chaotic dynamics.
In the following, we take the pair inertia function K to be an exponentially decaying function, first defined in \cite{classical_fractons}:

\begin{equation} \label{eq:K_nonloc}
    K(x) = \frac{1}{2} \, (\text{tanh}(\eta \, (x+a)) - \text{tanh}(\eta \, (x-a))).
\end{equation}

Lyapunov exponents measure the rate of separation of infinitesimally close trajectories, quantifying sensitivity to initial conditions. 
A positive non-zero Lyapunov exponent indicates chaos, as small differences in starting points lead to exponentially divergent outcomes.
From a simple counting argument, we show that all Lyapunov exponents for the three-fracton problem are zero. 
In the reduced coordinates, we have four degrees of freedom, hence, four possible exponents. 
Two of these exponents are zero as there are two directions along which exponents always disappear --- one along the trajectory and the second perpendicular to the manifold of constant energy. 
For fractons, there is a third zero exponent, from the emergent conservation of dipole moment in one of the clusters at late times. 
Finally, as we have a Hamiltonian system, all exponents come in pairs $(\lambda, -\lambda)$ \cite{Holger2003}. 
Thus, if just three exponents are zero, all four of them end up being zero. 
Hence, all Lyapunov exponents are zero: the three body case is completely non-chaotic.

\begin{figure}
\centering
\begin{subfigure}{8.6cm}
  \centering
  \caption{}
  \includegraphics[width=8.6cm]{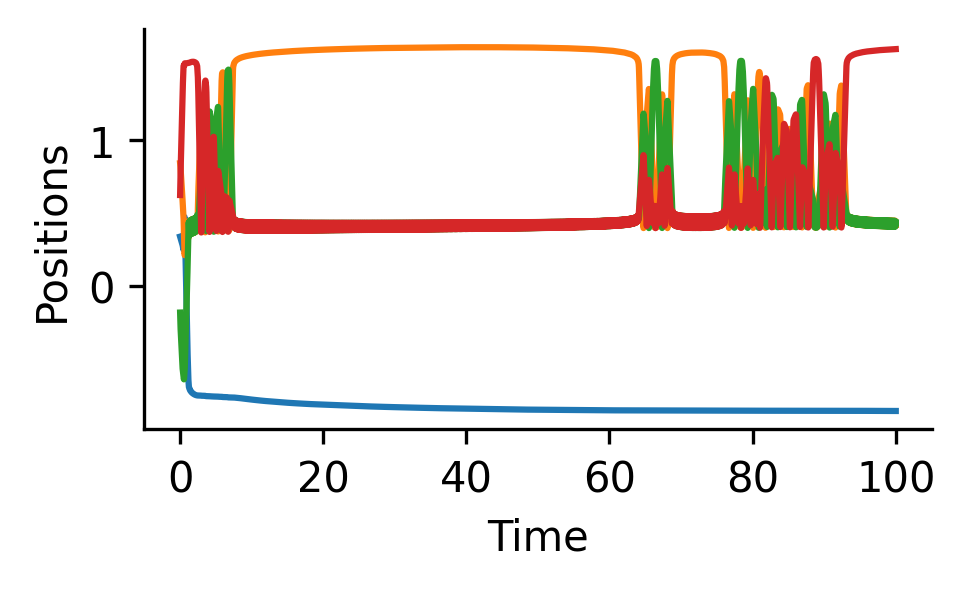}
  \label{fig:31lyapp1}
\end{subfigure}
\hfill
\begin{subfigure}{8.6cm}
  \centering
  \caption{}
  \includegraphics[width=8.6cm]{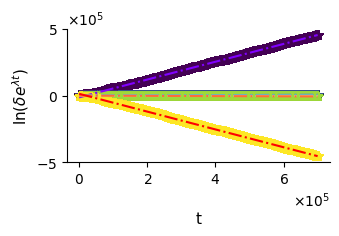}
  \label{fig:31lyape1}
\end{subfigure}
\caption{(a): Positions of the particles in a chaotic 3--1 trajectory ($\eta = 30$). (b): The Lyapunov exponents (given by the slopes) of the trajectory. }
\label{fig:31lyap1}
\end{figure}

Generalizing to four fractons, this guarantees only 4 of the 6 exponents to be zero, and indeed a pair of non-zero exponents is observed numerically\footnote{We note that fracton attractors are in position-\emph{velocity} space, yet we calculate Lyapunov exponents in conventional position-\emph{momentum} space. As fracton dynamics in phase space occurs over unbounded surfaces, the trajectories could exponentially diverge, whilst not being chaotic in a fuller sense. Our choice avoids cumbersome calculations, and the claims for chaos we present will carry over to position-velocity space.} for some 3--1 trajectories, e.g. in \cref{fig:31lyap1}.

Crucially, while one particle is at rest at all late times, there are brief intervals where another particle is also at rest. 
From our discussion in \ref{sec:31state}, this corresponds to the particle in the larger equilateral triangle (Note the triangle picture is only strictly valid for compact $K$. We apply it here by assuming the ``edges'' of $K$ are set by a sensible criteria, e.g.\ $K(x) = 0.1$). 
The region where a second particle is at rest corresponds to the part of the equilateral triangle extending outside the hexagon in \cref{fig:large_tri_hex}. 
These trajectories were \emph{not} observed in \cite{machian_fractons}, and are exclusive to non-compact pair inertia functions.

We compute the Lyapunov exponents for the four fracton trajectories, and find that 6 exponents are numerically close to zero, consistent with previous arguments, and the other two are of the form $(\lambda, -\lambda)$. 
In \cref{fig:31lyap1}, we show a typical trajectory and its Lyapunov exponents.
In reduced position space, late time trajectories are trapped in the reduced coordinate triangle.
In conventional systems with Lyapunov chaos, exponents for the same attractor are generally expected to be robust to different initial conditions (Multiplicative Ergodic Theorem \cite{skokos2010}).
However, this is not the case for fractons: different initial conditions cause the reduced coordinates triangle to be larger or smaller, so the late time states differ in how far the triangle extends beyond the hexagon. 

To summarize, by lifting the pair inertia function $K$ from a compact to a non-compact function, we introduce chaos.
This modifies the 3--1 late time state, introducing brief intervals with a momentarily frozen particle.
For more particles, $N>4$, we similarly expect the clustering at late times to vary, unlike the case of compact $K$, where clustering remains constant.
Therefore, we generally anticipate chaos for non-compact $K$.

\section{Janus Point and an Arrow of Time} \label{sec:arrow_of_time}

Zooming back out, we note that generic trajectories of our fracton system have the character that at late times they form the maximum number of clusters (``galaxies''). By time reversal invariance, this is also true if we run time backwards from such clustered states and look at large negative coordinate times. Somewhere in between, there is a time of maximum homogeneity (``the big bang'') which is our Janus point or central time, albeit one that can involve some degree of clustering as well.\footnote{In the extreme case of isolated fractons there is exactly the same degree of clustering at all times.} We now show that (i) there is a dynamical variable which we term ``complexity''\footnote{We have borrowed this term from Barbour et al \cite{Barbour2014} as there is a clear analogy between their work and ours. However, we should note that they are concerned with dimensionless measures of shape and hence their shape complexity is a ratio of two dynamical variables.} which captures clustering, and increases in both directions of coordinate time away from the Janus point, and (ii) a non-equilibrium Boltzmannian entropy which is correlated with the complexity.

\subsection{Complexity}

Through complexity, we aim to define a dynamical variable, which is a non-decreasing function away from a central time.

\begin{figure}[!h]
    \centering
    \begin{subfigure}{8.6cm}
        \centering
        \caption{}
        \includegraphics[width=\textwidth]{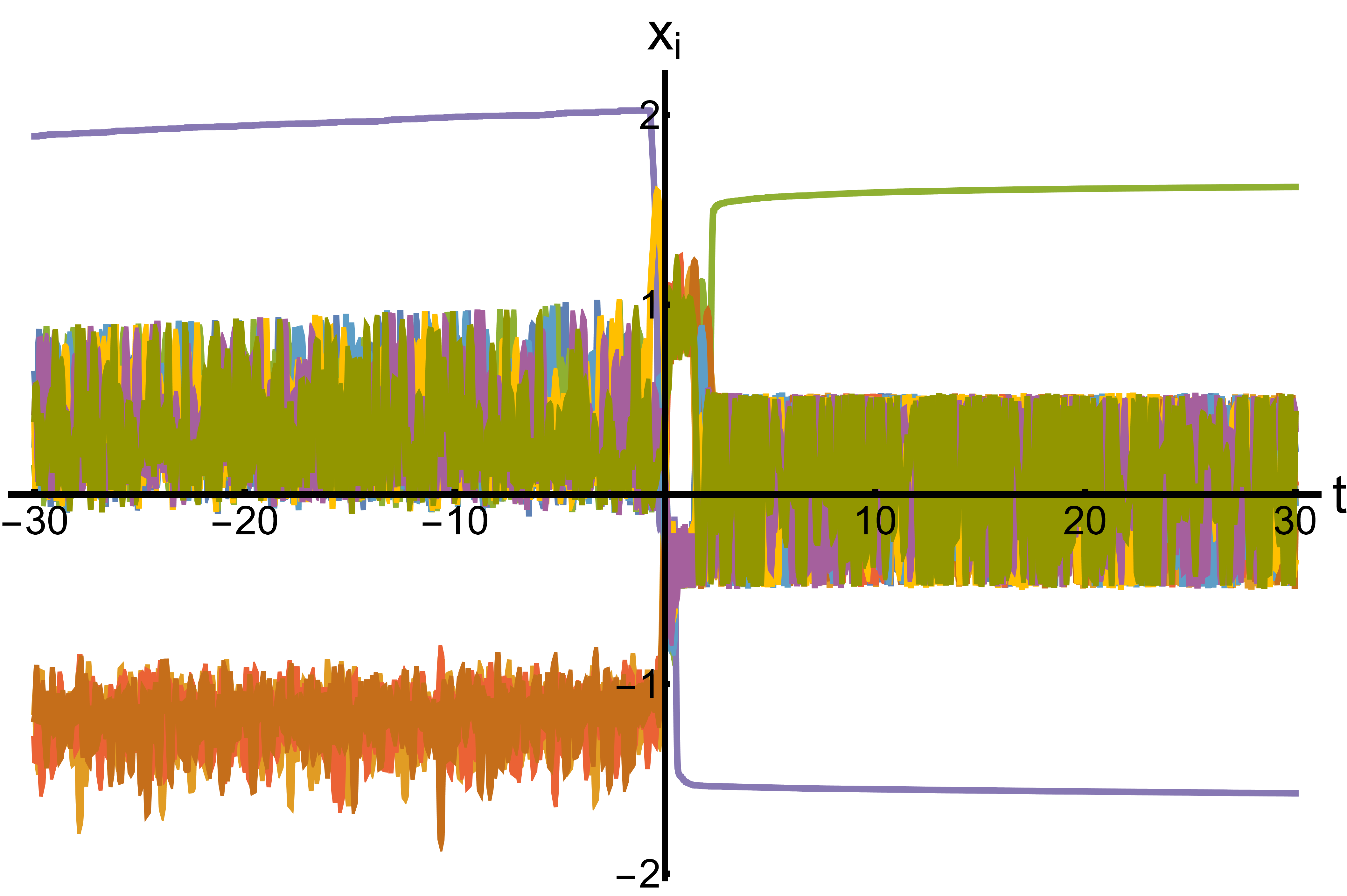} 
        \label{fig:pos_aot}
    \end{subfigure}
    \hfill
    \begin{subfigure}{8.6cm}
        \centering
        \caption{}
        \includegraphics[width=\textwidth]{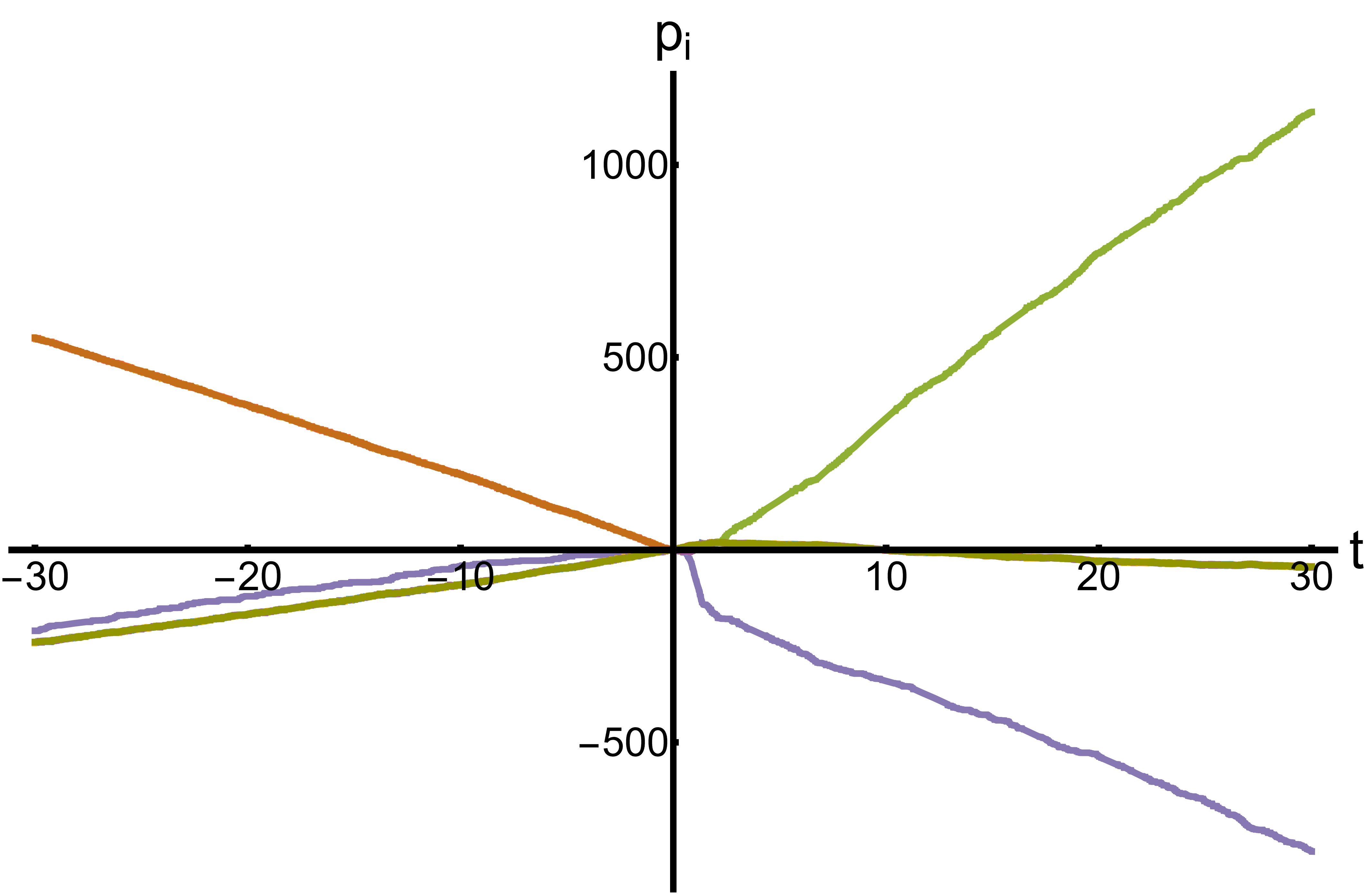} 
        \label{fig:mom_aot}
    \end{subfigure}
    \hfill
    \begin{subfigure}{8.6cm}
        \centering
        \caption{}
        \includegraphics[width=\textwidth]{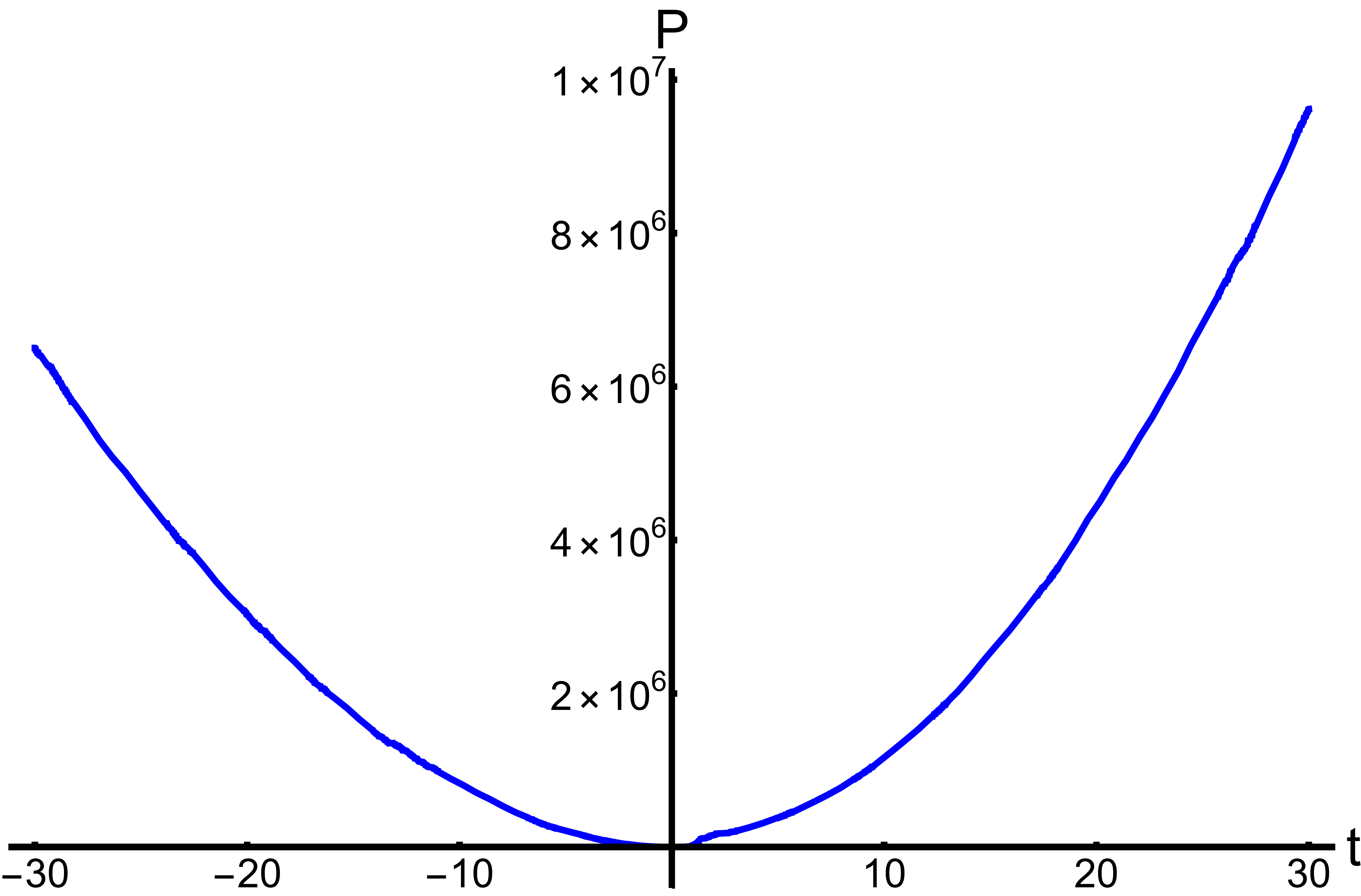} 
        \label{fig:p_aot}
    \end{subfigure}
    
    \caption{The time evolution of (a): positions, (b) momenta and (c) complexity, defined in \cref{eq:defp} for a system of 10 fractons. For the initial `big-bang' state, positions are randomly chosen within $[-0.5, 0.5]$ and momenta are chosen randomly within $[-1, 1]$. }
    \label{fig:aot_2}
\end{figure}

Once clustering has set in (\cref{fig:pos_aot}), positions are dense in a particular sub-space of position space. 
The clustering configuration is robust at late times, so a definition of complexity cannot come from positions alone.

The momenta (\cref{fig:mom_aot}) increase on either side of the initial ``big-bang'' configuration.
Therefore, the simplest complexity to consider would be sums of squares of momenta. 
However, each term must be invariant under translations of momenta, so we define complexity as:

\begin{equation} \label{eq:defp}
    P = \frac{1}{2} \sum_{i \neq j} \left(p_i - p_j\right)^2. 
\end{equation}

For what follows, we use the non-compact pair inertia function:
\begin{equation}
    K(x) = \frac{1}{2} \left( \text{tanh} \left(\eta(x+1)\right) - \text{tanh}\left(\eta(x-1)\right)\right).
\end{equation}

In \cref{fig:p_aot}, we plot the evolution of the complexity $P$ on either side of a big-bang initial configuration.
The complexity clearly increases on both sides of the ``big-bang''.
Our exploration of the space of trajectories and our detailed understanding of the dynamics in many cases strongly suggests that an increase of complexity from a central time is a generic feature of all non-stationary fractonic trajectories. 

\subsection{Entropy}

Thus far we have used a dynamical variable, the complexity, to identify Janus points. This is reminiscent of the shape complexity used in \cite{Barbour2014} which has a purely dynamical explanation for its temporal evolution. However the details are different as $P$ is not scale invariant and our systems have a characteristic scale set by the pair inertia function. So while we cannot rule out that we can construct a purely dynamical explanation for the Janus points, we look instead for an entropic explanation. In previous work we discussed the resemblance of the logic of fracton evolution to the phenomenon of ``order by disorder'' \cite{machian_fractons}. Here we follow \cite{Goldstein2016} and construct a non-equilibrium entropy connected to the complexity, which also shows increases about a central time, thus defining a thermodynamic arrow of time from the increase of entropy.

First we set up a toy model for late time states (e.g.\ for \cref{fig:aot_2}).
After clustering sets in at late times, the momenta increase linearly in time, neglecting small fluctuations.
In our toy model, we therefore take:
\begin{itemize}
\item Positions to be constrained within a fixed clustering configuration. 
\item Momenta to increase linearly in time. This ensures the clustering configuration is robust.
\end{itemize}

For a system of $N$ dipole conserving fractons, we have the following constants:

\begin{align}
    N x_{\text{cm}} &= \sum x_i, \\
    N p_{\text{cm}} &= \sum p_i.
\end{align}

These constants follow from invariance under translations $x_i \rightarrow x_i + \mu$ and $p_i \rightarrow p_i + \phi$. 
To simplify, we work with $x_\text{cm} = p_\text{cm} = 0$. 
The complexity then reduces to 

\begin{equation}
    P = \frac{1}{2}\sum_{i<j} (p_i - p_j)^2 = \frac{1}{4} \sum_i \sum_j (p_i - p_j)^2 = \frac{N}{2} \sum_i p_i^2,
    \label{eqn:complexity_gauge_0}
\end{equation}

We have assumed that once clustering has set in, on both sides of the Janus point:
\begin{equation}
    p_i = \alpha_i + \beta_i t,
\end{equation}
where $\alpha_i$ and $\beta_i$ are both constants, such the sum of both series from $i = 1$ to $N$ is zero. 
From now on we consider $t > 0$.
The process is identical for the other case, with different clustering, $\alpha_i$ and $\beta_i$.

We have that (dropping the factor of $N/2$):
\begin{equation}
    P \propto \left(\sum_{i=1}^N \left(\alpha_i\right)^2 \right) + 2\left(\sum_{i=1}^N \alpha_i \beta_i \right) t + \left(\sum_{i=1}^N \left(\beta_i\right)^2 \right) t^2.
\end{equation}

This can be written in the more instructive form

\begin{multline}
    P \propto \left(\sum_{i=1}^N \left(\beta_i\right)^2 \right) \left(t + \frac{\left(\sum_{i=1}^N \alpha_i \beta_i \right)}{\left(\sum_{i=1}^N \left(\beta_i\right)^2 \right)} \right)^2 +\\ \left(\sum_{i=1}^N \left(\alpha_i\right)^2 \right) - \frac{\left(\sum_{i=1}^N \alpha_i \beta_i \right)^2}{\left(\sum_{i=1}^N \left(\beta_i\right)^2 \right)}.
\end{multline}

This shows that for a time $t > -\frac{\sum_{i=1}^N \alpha_i^g \beta_i^g }{\sum_{i=1}^N \left(\beta_i^g\right)^2 } = t_+$, $P$ is monotonically increasing as a function of $t$. 
The same calculation can be repeated to show that $P$ is monotonically increasing as a function of $-t$ for times $t < t_-$. 
This suggests a picture where the complexity $P$ increases on either side of the big bang. 

In the same spirit as in \cite{Goldstein_Gibbs_v_boltzmann,Goldstein2016}, we now calculate the non-equilibrium Boltzmann entropy of this system. 
Roughly speaking, the Boltzmann entropy is defined by the log of the volume of the current ``macro-set'', being the set of all states that look similar to the current state.
By looking similar, we mean states of equal complexity $P$ and global conserved quantities (energy $E$, $x_{\mathrm{cm}}$ and $p_\mathrm{cm}$).
We then show our complexity increases with the entropy.

For a micro state, which is a set of $\{\{x_i\}, \{p_i\}\}$, we construct a surface in phase space ($\Gamma$), which has the same $x_\text{cm}, p_\text{cm}$ and energy $E$ as this micro state. 
$\Gamma$ is broken up into macro states $\Gamma_{[P, P+dP]}$, each with different values of the macro variable $P$.
The entropy is then a measure of the size of the current $\Gamma_{[P, P+dP]}$:

\begin{equation}
    S = k_B \text{ln} \left(\frac{\text{d}}{\text{d}P} \mathrm{Vol} (\Gamma_{[P, P+dP]} )\right).
\end{equation}

For our system of $N$ one-dimensional dipole conserving fractons, we have that the position and momentum phase space is $N-1$ dimensional, because we have constrained ourselves to $p_\text{cm} = x_\text{cm} = 0$. 
For simplicity, we shall assume that fractons are on a circle of length $l$ (identifying positions $-l/2$ and $l/2$). 
For position, we name this $N-1$ dimensional space $U_l$.
The position space is further restricted by the cluster configuration, which we now call $U_l^c$

Meanwhile, the space of momenta is isomorphic to $\mathbb{R}^{N-1}$
This is further constrained by energy.
For box pairwise inertia functions, $E$ becomes

\begin{equation}
    E = \frac{1}{2}\sum_{\substack{\text{cluster~}\\ c}} \sum_{\substack{i,j \in c\\i < j}} (p_i - p_j)^2,
\end{equation}
as we have restricted to a fixed cluster configuration in the toy model.
Note that there is no dependence on positions in $E$.
Hence, fixing $E$ only affects the phase space of momenta, reducing from $\mathbb{R}^{N-1}$ to the sub-space we denote $S_E^c$.
The total phase space is then $\Gamma_{[P, P+dP]} = U_l^c \times S_E^c$. 

The position space $U_l^c$ does not depend on the complexity $P$, so 
\begin{equation}
\frac{\text{d}}{\text{d}P} \mathrm{Vol}(U_l^c \times S_E^c) = \mathrm{Vol}(U_l^c) \frac{\text{d}}{\text{d}P} \mathrm{Vol}(S_E^c).
\end{equation}

Recalling \cref{eqn:complexity_gauge_0}, the complexity $P$ is defined by N/2 times the sum of squares of $p_i$: hence, the surface of constant $P$ is that of a hypersphere.
On $S_E^c$, consider the hyperspherical shell for radius squared $r^2$ between $P$ and $P + \text{d}P$.
Its volume is given by

\begin{align}
    \text{d}V' &\simeq f(N, S_E^c) \left(\sqrt{P}\right)^{N-3} \text{d}\sqrt{P} \\
    &\sim  f(N, S_E^c) \left(\sqrt{P}\right)^{N-4} \text{d}P,
\end{align}
where $f(N, S_E^c)$ depends on the clustering configuration, the energy and the number of fractons.
All together,
\begin{equation}
\frac{\text{d}}{\text{d}P} \mathrm{Vol} (\Gamma_{[P, P+dP]} )\simeq \text{Vol}(U_l^c) f(N, S_E^c) \left(\sqrt{P}\right)^{N-4}.
\end{equation}

Therefore, the Boltzmann entropy is given by 
\begin{equation}
    S \simeq k_B \left(\frac{N-4}{2}\right) \text{ln}(P) + \mathrm{const},
\end{equation}
where the constant does not depend on $P$.

Importantly, $S$ is an increasing function of $P$.
Complexity $P$, and hence the entropy $S$, is observed to be non-decreasing for generic trajectories (e.g.\ \cref{fig:aot_2}), in accordance with the second law.
The construction of this entropy produces a thermodynamic arrow of time, defined by the direction of increasing entropy about the Janus point.

\section{In closing}

Classical fractons turn out to be an exceptionally interesting dynamical system. 
Symmetries and locality require the minimal, Machian, dynamics to be non-linear but the additional conservation laws that restrict the mobility of the particles also lead to a breakdown of global ergodicity. 
In previous work we showed that this ergodicity breaking goes along with ``violations'' of the Liouville and Hohenberg-Mermin-Wagner-Coleman theorems, but more precisely of common beliefs about what these theorems imply about the existence of attractors in Hamiltonian systems and about the breaking of continuous symmetries in low dimensions.

In this paper we have advanced our understanding of these systems in two directions. 
First, we have presented evidence that the breakdown of global ergodicity coexists with local chaos (or at least local ergodicity) in the clustered spatial organization of fracton attractors. 
Second, we have shown that generic trajectories of these systems exhibit Janus points---times of maximal homogeneity---which then define two arrows of time leading in either direction of coordinate time. 

Somewhat surprisingly, many of the above features of fractons have a family resemblance to some ideas in the cosmology literature. 
Fracton dynamics is Machian, a term that makes reference to Mach's ideas on the origin of inertia. Inflationary models are the other example we have found of Hamiltonian systems that exhibit attractor dynamics in defiance of Liouville's theorem \cite{PhysRevD.88.083518}. Janus points were discovered in Cosmology in an attempt to provide a theory of the arrow of time without appealing to a past hypothesis. The portability of ideas across vast ranges of physical scale remains one of the truly remarkable features of Physics!

It was pointed out earlier in Refs.~\cite{classical_fractons,machian_fractons} that the ergodicity breaking in our classical systems resembles the strong Hilbert space fragmented phase of lattice fractons at low particle density. In future work, it would be illuminating to see if aspects of this work, such as local chaos and the emergence of an arrow of time are also observed there. Also interesting would be to see how these phenomena are affected as ergodicity is partially restored at high densities when the system transitions to the weakly Hilbert space fragmented phase. \cite{classenhowes2025universalfreezingtransitionsdipoleconserving,MorninstarKhemaniHuse_PhysRevB.101.214205,Pozederac_PhysRevB.107.045137}. More generally, it would be particularly interesting to investigate whether some of these fractonic features, especially the Janus point phenomenon, show up in other non-equilibrium systems of interest to many-body physicists.

\medskip 
\noindent\emph{Acknowledgments}: We thank Julian Barbour for a discussion of his work and Marc Warner for introducing SLS to the former. The work of A.P. was supported by the European Research Council under the European Union Horizon 2020 Research and Innovation Programme, Grant Agreement No. 804213-TMCS  and the Engineering and Physical Sciences Research Council, Grant number EP/S020527/1. The work of S.L.S. was supported by a Leverhulme Trust International Professorship, Grant Number LIP-202-014. For the purpose of Open Access, the author has applied a CC BY public copyright license to any Author Accepted Manuscript version arising from this submission.

\appendix

\section{Three fracton problem} \label{appendix:three_fracton}

\noindent We present three fracton trajectories for all possible initial conditions. The symmetry in the hexagonal picture is exploited, and hence results that apply to one of the 6 sub-regions in 2 or 1 apply to all the sub-regions in 2 or 1. \\

\vspace{5mm}

\textbf{Trajectory starting out in region 3}

\vspace{5mm}

\noindent For our trajectory beginning in region 3, we suppose that the trajectory goes from 3 to $2_a$, and from $2_a$ to $1_b$. In region 3, and in $2_a$, we necessarily require that $\dot q_1^{\left[3\right]} > 0$, given that in region 3, $(\dot q_1^{\left[3\right]}, \dot q_2^{\left[3\right]}) = (3\pi_1^{\left[3\right]}, 3\pi_2^{\left[3\right]})$, our necessary condition to enter region $2_a$ translates to $\pi_1^{\left[3\right]} > 0$. \\

\noindent From \cite{classical_fractons}, we have that $\sqrt{3} \pi_1^{\left[3\right]} = \pi_1^{\left[2_a\right]}, \pi_2^{\left[3\right]} = \pi_2^{\left[2_a\right]}$. We now suppose that our trajectory strikes the $2_{a} - 1_b$ interface. The discontinuous change in momentum is governed by the two equations, 
\begin{equation}
    \sqrt{3} \pi_1^{\left[1_b\right]} - \pi_2^{\left[1_b\right]} = \sqrt{3} \pi_1^{\left[2_a\right]} - \pi_2^{\left[2_a\right]},
\end{equation}
\begin{equation}
    \frac{(\sqrt{3} \pi_2^{\left[1_b\right]} - \pi_1^{\left[1_b\right]})^2}{4} = \frac{(\sqrt{3} \pi_2^{\left[2_a\right]} - \pi_1^{\left[2_a\right]})^2}{4} + \frac{(\sqrt{3} \pi_2^{\left[2_a\right]} + \pi_1^{\left[2_a\right]})^2}{4},
\end{equation}

\noindent where the first equation arises from eliminating the undefined $B'\left(\frac{\sqrt{3}q_2 + q_1}{\sqrt{2}}\right)$ from the equations for $\dot \pi_1$ and $\dot \pi_2$. From these equations, we will obtain two pairs of solutions, and from these two pairs, the pair that gives the solution where the trajectory moves away from the $2_a - 1_b$ wall after striking the interface is chosen. This yields the solution
\begin{equation} \label{eq:1bfrom2a1}
    \pi_1^{\left[1_b\right]} = \frac{3}{2} \pi_1^{\left[2_a\right]} - \frac{\sqrt{3}}{2} \pi_2^{\left[2_a\right]} + \frac{1}{\sqrt{2}} \sqrt{(\pi_1^{\left[2_a\right]})^2 + 3 (\pi_2^{\left[2_a\right]})^2},
\end{equation}
\begin{equation} \label{eq:1bfrom2a2}
    \pi_2^{\left[1_b\right]} = \frac{\sqrt{3}}{2} \pi_1^{\left[2_a\right]} - \frac{1}{2} \pi_2^{\left[2_a\right]} + \frac{\sqrt{3}}{\sqrt{2}} \sqrt{(\pi_1^{\left[2_a\right]})^2 + 3 (\pi_2^{\left[2_a\right]})^2},
\end{equation}

\noindent for momenta in $1_b$. The trajectories then strike the $1_b - 2_f$ wall, and the discontinuous change in momentum is described by the two equations

\begin{equation}
    \pi_2^{\left[2_f\right]} = \pi_2^{\left[1_b\right]},
\end{equation}
\begin{equation}
    (\pi_1^{\left[2_f\right]})^2 + \frac{(\sqrt{3}\pi_2^{\left[2_f\right]} - \pi_1^{\left[2_f\right]})^2}{4} = \frac{(\sqrt{3}\pi_2^{\left[1_b\right]} - \pi_1^{\left[1_b\right]})^2}{4}.
\end{equation}

\noindent These equations, combined with the condition that the trajectory moves away from the $1_b - 2_f$ boundary, i.e., $\dot q_1 < 0$ yield 

\begin{equation} \label{eq:2ffrom1b1}
    \pi_1^{\left[2_f\right]} = \frac{2\sqrt{3}\pi_2^{\left[1_b\right]} - \sqrt{12(\pi_2^{\left[1_b\right]})^2 + 20 (\pi_1^{\left[1_b\right]})^2 - 40\sqrt{3}\pi_1^{\left[1_b\right]} \pi_2^{\left[1_b\right]}}}{10},
\end{equation}
\begin{equation} \label{eq:2ffrom1b2}
    \pi_2^{\left[2_f\right]} = \pi_2^{\left[1_b\right]}.
\end{equation}

\noindent We create a Mathematica RegionPlot with the condition $12(\pi_2^{\left[1_b\right]})^2 + 20 (\pi_1^{\left[1_b\right]})^2 - 40\sqrt{3}\pi_1^{\left[1_b\right]} \pi_2^{\left[1_b\right]} > 0$, which is just checking if the solution is real or not, and in any arbitrarily large region constrained by $\pi_1^{\left[3\right]}>0$, the discriminant, given in terms of $\pi_1^{\left[3\right]}$ and $\pi_2^{\left[3\right]}$ is negative, thus we conclude that the trajectory does not enter region $2_f$ and is reflected back into region $1_b$. We shall now refer to the momentum before reflection as $\pi_{i<}^{\left[1_b\right]}$, and the momentum after reflection as $\pi_{i>}^{\left[1_b\right]}$. The equations governing this discontinuous change in momentum are
\begin{equation}
    \pi_{2>}^{\left[1_b\right]} = \pi_{2<}^{\left[1_b\right]},
\end{equation}
\begin{equation}
    \sqrt{3}\pi_{2>}^{\left[1_b\right]} - \pi_{1>}^{\left[1_b\right]} = \pi_{1<}^{\left[1_b\right]} - \sqrt{3}\pi_{2<}^{\left[1_b\right]}.
\end{equation}

\noindent Both these constraints are the same constraints as the ones for trajectories going into $2_f$ from $1_b$. Our trajectory now approaches $2_a$ from $1_b$. From \cite{classical_fractons}, we have that the momenta are given by

\begin{multline}
    \pi_1^{\left[2_a\right]} = \frac{9}{10} \pi_1^{\left[1_b\right]} - \frac{3\sqrt{3}}{10} \pi_2^{\left[1_b\right]} \\
    - \frac{1}{5} \sqrt{3 (\pi_2^{\left[1_b\right]})^2 - (\pi_1^{\left[1_b\right]})^2 - \sqrt{3} \pi_1^{\left[1_b\right]}\pi_2^{\left[1_b\right]}},
\end{multline}

\begin{multline}
    \pi_2^{\left[2_a\right]} = -\frac{\sqrt{3}}{10} \pi_1^{\left[1_b\right]} + \frac{1}{10} \pi_2^{\left[1_b\right]} \\
    - \frac{\sqrt{3}}{5} \sqrt{3 (\pi_2^{\left[1_b\right]})^2 - (\pi_1^{\left[1_b\right]})^2 - \sqrt{3} \pi_1^{\left[1_b\right]}\pi_2^{\left[1_b\right]}}.
\end{multline}

\noindent For these equations, the momenta $\pi_1^{\left[1_b\right]}$ and $\pi_2^{\left[1_b\right]}$ are given by $\pi_{1>}^{\left[1_b\right]}$ and $\pi_{2>}^{\left[1_b\right]}$, and creating a Mathematica RegionPlot with the condition $3(\pi_{2>}^{\left[1_b\right]})^2 -(\pi_{1>}^{\left[1_b\right]})^2 - \sqrt{3} (\pi_{1>}^{\left[1_b\right]}) (\pi_{2>}^{\left[1_b\right]}) > 0$ in terms of $\pi_1^{\left[3\right]}$, we find that any arbitrarily large region, independent of it obeying the constraint $\pi_1^{\left[3\right]} > 0$, does not have any point satisfying the constraint, so the trajectory bounces off the $1_b- 2_f$ wall also.

\noindent It is easy to prove that if we consider the trajectory going towards the $1_b - 2_f$ wall, after $n$ pairs of reflections from the $1_b - 2_f$ wall and $1_b - 2_a$ wall, the momentum going into the $1_b - 2_f$ wall for the $n^{\text{th}}$ time is 
\begin{align*}
    \pi_{1<n}^{\left[1_b\right]} = (1-3n) \pi_{1<}^{\left[1_b\right]} + 3\sqrt{3}n\pi_{2<}^{\left[1_b\right]}, \\
    \pi_{2<n}^{\left[1_b\right]} = -\sqrt{3}n \pi_{1<}^{\left[1_b\right]} + (1+3n)\pi_{2<}^{\left[1_b\right]},
\end{align*} 

and in a similar construction, the momentum going into the $1_b - 2_a$ wall for the $n^{\text{th}}$ time is 
\begin{align*}
    & \pi_{1>n}^{\left[1_b\right]} = (1+3n) \pi_{1>}^{\left[1_b\right]} - 3\sqrt{3}n\pi_{2>n}^{\left[1_b\right]}, \\
    & \pi_{2>n}^{\left[1_b\right]} = \sqrt{3}n \pi_{1>}^{\left[1_b\right]} + (1-3n)\pi_{2>}^{\left[1_b\right]},
\end{align*}.

\noindent Constructing similar Mathematica RegionPlots for $n>0$ tells us that as soon as the trajectory strikes the $1_b - 2_f$ or $1_b - 2_a$ wall, it is reflected back into $1_b$. We thus conclude that any trajectory that enters $2_a$ from 3 and goes to $1_b$ from $2_a$ gets trapped in $1_b$ (\cref{fig:3fractrajectory}a). From the symmetry of the problem the trajectory gets trapped in $1_c$ should the trajectory enter $1_c$ from $2_a$, and this result straightforwardly generalizes to trajectories that enter other sub-regions, $2_b - 2_f$, from region 3.

\vspace{5mm}

\textbf{Trajectory starting out in sub-region $2_a$}

\vspace{5mm}

\noindent If our trajectory starting out in $2_a$ strikes the $2_a - 3$ boundary, it will always enter 3. After it has entered 3, we know exactly what happens to the trajectory. We suppose that the trajectory starting out in sub-region $2_a$ goes toward the $2_a - 1_b$ boundary. All such trajectories enter $1_b$ with momenta given by (\ref{eq:1bfrom2a1}) and (\ref{eq:1bfrom2a2}), and all these trajectories move toward the $1_b - 2_f$ wall. Some of these trajectories get reflected back into $1_b$ from there, while the others pass into $2_f$. All trajectories that get reflected back into $1_b$ from the $1_b - 2_f$ wall are trapped in the region $1_b$ with the momentum increasing as detailed in the previous section. In hindsight of the previous section, this does not come as a surprise. While reflection from the $1_b - 2_f$ wall is not true for all possible initial momenta (it is true for all the trajectories that enter $2_a$ from 3, but the space of trajectories that enter $2_a$ from 3 is distinct from the space of trajectories entering $1_b$ from $2_a$), after a reflection from $1_b - 2_f$ wall into $1_b$, all trajectories that started in 3 necessarily get reflected from the $1_b - 2_a$ wall. This is also true for trajectories originating in $2_a$ because of the fact that we only need to change $\pi_1^{\left[3\right]}$ to $\pi_1^{\left[2_a\right]}/\sqrt{3}$ and $\pi_2^{\left[3\right]}$ to $\pi_2^{\left[2_a\right]}$ in the plots, which is fine because the plots show reflection for arbitrarily large ranges of momenta. \\

\noindent We now suppose that the trajectory enters $2_f$. To strike the $2_f - 3$ wall, a necessary, but not sufficient condition is that the trajectory should have a positive component of motion perpendicular to the direction of the wall, towards the wall. This means that we must have $\sqrt{3} \dot q_2 + \dot q_1 < 0$, which in sub-region $2_f$ translates to $\sqrt{3}\pi_2^{\left[2_f\right]} + \pi_1^{\left[2_f\right]} < 0$. Constructing a Mathematica RegionPlot for $\sqrt{3}\pi_2^{\left[2_f\right]} + \pi_1^{\left[2_f\right]} < 0$ in terms of $\pi_1^{\left[2_a\right]}$ and $\pi_2^{\left[2_a\right]}$, we find that for an arbitrarily large region in $\pi_1^{\left[2_a\right]}$ and $\pi_2^{\left[2_a\right]}$ space, this condition does not hold true. Given that entering $2_f$ from $1_b$ would mean that the trajectory is moving away from the $2_f - 1_b$ wall, and we have shown that the trajectory is also moving away from the $2_f - 3$ wall, all trajectories entering $2_f$ from $2_a$ via $1_b$ will necessarily advance into $1_a$.

\noindent To describe the discontinuous change of momentum going in from $2_f$ to $1_a$, we use the equations 

\begin{equation}
    \sqrt{3} \pi_1^{\left[1_a\right]} + \pi_2^{\left[1_a\right]} = \sqrt{3} \pi_1^{\left[2_f\right]} + \pi_2^{\left[2_f\right]},
\end{equation}
\begin{equation}
    (\pi_1^{\left[1_a\right]})^2 = \frac{(\sqrt{3} \pi_2^{\left[2_f\right]} - \pi_1^{\left[2_f\right]})^2}{4} + (\pi_1^{\left[2_f\right]})^2,
\end{equation}

\noindent coupled with the condition that $\pi_1^{\left[1_a\right]} < 0$ to get
\begin{equation}
    \pi_1^{\left[1_a\right]} = - \sqrt{\frac{(\sqrt{3} \pi_2^{\left[2_f\right]} - \pi_1^{\left[2_f\right]})^2}{4} + (\pi_1^{\left[2_f\right]})^2},
\end{equation}
\begin{equation}
    \pi_2^{\left[1_a\right]} = \sqrt{3} \pi_1^{\left[2_f\right]} + \pi_2^{\left[2_f\right]} + \sqrt{3} \sqrt{\frac{(\sqrt{3} \pi_2^{\left[2_f\right]} - \pi_1^{\left[2_f\right]})^2}{4} + (\pi_1^{\left[2_f\right]})^2}.
 \end{equation}

\noindent All these trajectories necessarily strike the $1_a - 2_e$ wall, for which the momenta are determined using 
\begin{equation}
    \frac{(\sqrt{3} \pi_2^{\left[2_e\right]} + \pi_1^{\left[2_e\right]})^2}{4} + (\pi_1^{\left[2_e\right]})^2 = (\pi_1^{\left[1_a\right]})^2 ,
\end{equation}

\begin{equation}
    \sqrt{3} \pi_1^{\left[2_e\right]} - \pi_2^{\left[2_e\right]} = \sqrt{3} \pi_1^{\left[1_a\right]} - \pi_2^{\left[1_a\right]}.
\end{equation}

\noindent The discriminant of the quadratic to obtain $\pi_1^{\left[2_e\right]}$ is $D_1 = -48 (\pi_2^{\left[1_a\right]})^2 + 176(\pi_1^{\left[1_a\right]})^2 + 96 \sqrt{3} \pi_1^{\left[1_a\right]} \pi_2^{\left[1_a\right]}$. If we create the Mathematica RegionPlot for $D_1 > 0$ in terms of the variables $\pi_1^{\left[2_a\right]}$ and $\pi_2^{\left[2_a\right]}$, we find that for an arbitrarily large region in $\pi_1^{\left[2_a\right]}$ and $\pi_2^{\left[2_a\right]}$ space, this condition does not hold true for any point, so the trajectory reflects back into $1_a$. It then strikes the $1_a - 2_f$ interface, for which the new momentum after reflection is given by the relations

\begin{equation} \label{eq:greatintermsofless1a1}
    \pi_{1>}^{\left[1_a\right]} = - \pi_{1<}^{\left[1_a\right]},
\end{equation}

\begin{equation} \label{eq:greatintermsofless1a2}
    \pi_{2>}^{\left[1_a\right]} =  2\sqrt{3}\pi_{1<}^{\left[1_a\right]} + \pi_{2<}^{\left[1_a\right]}.
\end{equation}

\noindent The discriminant for calculating $\pi_1^{\left[2_f\right]}$ is $D_2 = -48 (\pi_{2>}^{\left[1_a\right]})^2 + 176(\pi_{1>}^{\left[1_a\right]})^2 - 96 \sqrt{3} \pi_{1>}^{\left[1_a\right]} \pi_{2>}^{\left[1_a\right]}$, and repeating the process, we find that the trajectory gets reflected from the $1_a - 2_f$ wall as well.

\noindent It is easy to verify that after $n$ pairs of reflections from the $1_a - 2_f$ wall and $1_a - 2_e$ wall, the momentum going into the $1_a - 2_e$ wall for the $n^{\text{th}}$ time is 
\begin{align} \label{eq:lessn1a}
    & \pi_{1<n}^{\left[1_a\right]} = \pi_{1<}^{\left[1_a\right]}, \nonumber \\
    & \pi_{2<n}^{\left[1_a\right]} = -4\sqrt{3}n \pi_{1<}^{\left[1_a\right]} + \pi_{2<}^{\left[1_a\right]},
\end{align}

\noindent and the momentum after $n$ pairs of reflections from the $1_a - 2_e$ wall and $1_a - 2_f$ wall, the momentum going into the $1_a - 2_f$ wall for the $n^{\text{th}}$ time is 
\begin{align} \label{eq:greatn1a}
    & \pi_{1>n}^{\left[1_a\right]} = \pi_{1>}^{\left[1_a\right]}, \nonumber \\
    & \pi_{2>n}^{\left[1_a\right]} = 4\sqrt{3}n \pi_{1>}^{\left[1_a\right]} + \pi_{2>}^{\left[1_a\right]}.
\end{align}

\noindent Using the same method, the trajectories get reflected from the $1_a - 2_e$ and $1_a - 2_f$ boundaries to remain trapped in $1_a$. Thus, for trajectories that start out in $2_a$ moving towards $1_b$, they can either get trapped in $1_b$ (\cref{fig:3fractrajectory}a) or end up trapped in $1_a$ (\cref{fig:3fractrajectory}b). Again, this generalizes to moving towards $1_c$ from $2_a$, and even to starting in different sub-regions $2_b - 2_f$.

\vspace{5mm}

\textbf{Trajectory starting out in sub-region $1_a$}

\vspace{5mm}

\noindent This is perhaps the most interesting set of initial conditions. From the last section, we know that the discriminant for calculating $\pi_1^{\left[2_e\right]}$ is $D_1 = -48 (\pi_2^{\left[1_a\right]})^2 + 176(\pi_1^{\left[1_a\right]})^2 + 96 \sqrt{3} \pi_1^{\left[1_a\right]} \pi_2^{\left[1_a\right]}$, and that for calculating $\pi_1^{\left[2_f\right]}$ is $D_2 = -48 (\pi_{2>}^{\left[1_a\right]})^2 + 176(\pi_{1>}^{\left[1_a\right]})^2 - 96 \sqrt{3} \pi_{1>}^{\left[1_a\right]} \pi_{2>}^{\left[1_a\right]}$. We also know from (\ref{eq:greatintermsofless1a1}) and (\ref{eq:greatintermsofless1a2}), the relation between $\pi_{i<}^{\left[1_a\right]}$ and $\pi_{i>}^{\left[1_a\right]}$, and from (\ref{eq:lessn1a}) and (\ref{eq:greatn1a}), the momentum after $n$ reflections. We suppose that when we start from $1_a$, the initial momentum, $\pi_1^{\left[1_a\right]} < 0$, such that the trajectory strikes the $1_a - 2_e$ wall, thus, if we assume reflection, in terms of the initial momentum, the discriminant $D_2$ becomes
\begin{equation}
    D_2 = -16 \, \left(61 \left(\pi_1^{\left[1a\right]}\right)^2 - 18\sqrt{3} \pi_1^{\left[1a\right]} \pi_2^{\left[1a\right]} + 3 \left(\pi_1^{\left[1a\right]}\right)^2\right).
\end{equation}

\noindent After $n$ pairs of reflections, we have that the discriminant $D_1^n$ and $D_2^n$ are given by

\begin{align}
    D_1^n &= -16 \, \big((-11+72n+144n^2)\left(\pi_1^{\left[1a\right]}\right)^2 \nonumber \\
    &\quad - 6\sqrt{3} (1 + 4n)\pi_1^{\left[1a\right]}\pi_2^{\left[1a\right]} + 3 \left(\pi_2^{\left[1a\right]}\right)^2\big), \nonumber \\
    D_2^n &= -16 \, \big((61+216n+144n^2)\left(\pi_1^{\left[1a\right]}\right)^2 \nonumber \\
    &\quad - 6\sqrt{3} (3 + 4n)\pi_1^{\left[1a\right]}\pi_2^{\left[1a\right]} + 3 \left(\pi_2^{\left[1a\right]}\right)^2\big). \label{eq:D2n}
\end{align}

\noindent The region in phase space, where $D_1^n < 0$ is given by $\pi_2^{\left[1a\right]}/\pi_1^{\left[1a\right]} \in S_1^n$, and the region where $D_2^n < 0$ is given by $\pi_2^{\left[1a\right]}/\pi_1^{\left[1a\right]} \in S_2^n$, where sets $S_1^n$ and $S_2^n$ are given, $\forall \, n \in \mathbb{Z}^+ \cup {0}$ by
\begin{align}
    S_1^n = \left(-\infty, (1+4n)\sqrt{3} - \frac{2}{3} \sqrt{15}\right) \nonumber \\
    \cup \left((1+4n)\sqrt{3} + \frac{2}{3} \sqrt{15}, \infty\right), \nonumber \\
    S_2^n = \left(-\infty, (3+4n)\sqrt{3} - \frac{2}{3} \sqrt{15}\right) \nonumber \\
    \cup \left((3+4n)\sqrt{3} + \frac{2}{3} \sqrt{15}, \infty\right).
\end{align}

\noindent Therefore, we have that for the trajectory to exit as soon as it strikes the $2_e$ wall for the first time,
\begin{align}
    \pi_2^{\left[1a\right]}/\pi_1^{\left[1a\right]} \in \Bar{S_1^0},
\end{align}

\noindent where, for our sets $S_i^j$, $\Bar{S_i^j}$ is defined by $\Bar{S_i^j} \cap S_i^j = \emptyset$ and $\Bar{S_i^j} \cup S_i^j = \mathbb{R}$. This means that for the aforementioned case,
\begin{align}
    \pi_2^{\left[1a\right]}/\pi_1^{\left[1a\right]} \in \left[\sqrt{3} - \frac{2}{3} \sqrt{15} , \sqrt{3} + \frac{2}{3} \sqrt{15}\right].
\end{align}

\noindent Similarly, for one reflection by $2_e$ and then transmission as soon as it hits the $2_f$ wall, 
\begin{equation}
    \pi_2^{\left[1a\right]}/\pi_1^{\left[1a\right]} \in S_1^0 \cap \Bar{S_2^0} = \left(\sqrt{3} + \frac{2}{3} \sqrt{15} , 3\sqrt{3} + \frac{2}{3} \sqrt{15}\right].
\end{equation}

\noindent The pattern starts becoming more clear when we consider one pair of reflections, each by the $2_e$ and $2_f$ wall, followed by transmission through the $2_e$ wall. For this
\begin{equation}
    \pi_2^{\left[1a\right]}/\pi_1^{\left[1a\right]} \in S_1^0 \cap S_2^0 \cap \Bar{S_1^1} = \left(3\sqrt{3} + \frac{2}{3} \sqrt{15} , 5\sqrt{3} + \frac{2}{3} \sqrt{15}\right].
\end{equation}

\noindent Therefore, for a total of $m$ reflections, $(m>1)$ before exiting $1_a$, where $m$ is the sum of the number of reflections by the $1_a - 2_e$ and $1_a - 2_f$ walls, we have that
\begin{equation}
    \pi_2^{\left[1a\right]}/\pi_1^{\left[1a\right]} \in \left((2m-1)\sqrt{3} + \frac{2}{3} \sqrt{15} , (2m+1)\sqrt{3} + \frac{2}{3} \sqrt{15}\right].
\end{equation}

\noindent This means that we can choose our initial momentum such that after any finite number of reflections, the trajectory exits the region $1_a$. This is in complete contrast to what we observed when we started from the sub-regions of 2 and the region 3, where once the particle has been reflected once by any wall of any sub-regions of 1, it is trapped in that region.

\noindent For being trapped in region $1_a$, the condition is
\begin{equation}
    \pi_2^{\left[1a\right]}/\pi_1^{\left[1a\right]} \in \bigcap\limits_{i=0}^{\infty} \left(S_1^i \cap S_2^i\right) = \left(-\infty , \sqrt{3} - \frac{2}{3} \sqrt{15}\right).
\end{equation}

\noindent The only possibility that is not captured by these conditions is if $\pi_1^{\left[1_a\right]} = 0$, in which case we have that the trajectory does not move in $(q_1, q_2)$ phase space, the initial momentum of the trajectory is its momentum at all times.  \\

\noindent After the trajectory has exited the region $1_a$, it enters one of the sub-regions of $2$, for which we know what happens to the trajectory.

\noindent A very interesting extension of this problem is asking what happens to the trajectory in reverse time. We shall pick our starting region to be $1_a$ for simplicity. It can be shown that the trajectory exists the sub-region when it first strikes the $1_a - 2_f$ boundary in reverse time (for $\pi_1^{[1_a]} < 0$) when 

\begin{equation}
    \pi_2^{\left[1a\right]}/\pi_1^{\left[1a\right]} \in \left[-\sqrt{3} - \frac{2}{3} \sqrt{15} , -\sqrt{3} + \frac{2}{3} \sqrt{15}\right].
\end{equation}

\noindent We can also shown that after striking the $1_a - 2_f$ boundary once, reflecting back into into $1_a$ the exits via the $1_a - 2_e$ boundary if

\begin{equation}
    \pi_2^{\left[1a\right]}/\pi_1^{\left[1a\right]} \in \left[-3\sqrt{3} - \frac{2}{3} \sqrt{15} , -\sqrt{3} - \frac{2}{3} \sqrt{15}\right).
\end{equation}

\noindent Similar to the previous case, we have that the trajectory would leave $1_a$ in reverse time after $m$ total bounces against the $1_a - 2_f$ and $1_a - 2_e$ wall if

\begin{equation}
    \frac{\pi_2^{\left[1a\right]}}{\pi_1^{\left[1a\right]}} \in \left[-(2m+1)\sqrt{3} - \frac{2}{3} \sqrt{15} ,
    -(2m-1)\sqrt{3} - \frac{2}{3} \sqrt{15}\right).
\end{equation}

\noindent There is an overlap between ranges of the ratio of the reduced momenta where the trajectory escapes in reverse time and the trajectory escapes in forward time -

\begin{equation} \label{eq:escforback}
    \pi_2^{\left[1a\right]}/\pi_1^{\left[1a\right]} \in \left[-\sqrt{3} - \frac{2}{3} \sqrt{15} , -\sqrt{3} + \frac{2}{3} \sqrt{15}\right].
\end{equation}

\noindent This case corresponds to trajectories like the one in \cref{fig:3fractrajectory}b. Should we look at the system when it was in $1_b$ in this case, we shall see that in forward time, it escapes as soon as it strikes $1_b - 2_f$ interface and in reversed time, it escapes as soon as it strikes the $1_b - 2_a$ interface - precisely what we see in (\ref{eq:escforback}).
\vspace{5mm}

\bibliography{references}

\end{document}